\documentclass[structabstract]{aa} 
\usepackage{natbib,amssymb,graphicx,graphics}
\usepackage{float}
\usepackage{graphicx}
\usepackage{txfonts}

\def\L'{m$_{L'}$\ }

\def\aap{A\&A\ }

\def\apjl{ApJL\ }

\def\apjs{ApJS\ }

\DeclareGraphicsExtensions{.jpg, .png, .bmp, .eps, .pdf, .tiff}

\begin{document} 

   \title{New constraints on the formation and settling of dust in the atmospheres of young M and L dwarfs\thanks{Based on observations made with ESO telescopes at the Paranal
Observatory under programs 085.C-0676 and 290.C-5145.}}

   \author{E. Manjavacas\inst{1}
      \and M. Bonnefoy\inst{1,2}
	  \and J. E. Schlieder\inst{1}
	  \and F. Allard\inst{3}
	  \and P. Rojo\inst{4}
	  \and B. Goldman\inst{1}
	  \and G. Chauvin\inst{2}
	  \and D. Homeier\inst{3}
	  \and N. Lodieu\inst{5, 6}
	  \and T. Henning\inst{1}
          }

	\institute{Max Planck Institute f\"ur Astronomie. K\"onigstuhl, 17. D-69117 Heidelberg, Germany\\
	\email{manjavacas@mpia.de}
	\and UJF-Grenoble1/CNRS-INSU,Institut de Plan\'{e}tologie et d'Astrophysique de Grenoble (IPAG) UMR5274, Grenoble 38041, France
	\and CRAL-ENS, 46, All\'{e}e d'Italie , 69364 Lyon Cedex 07, France
	\and Departamento de Astronomia, Universidad de Chile, Casilla 36-D, Santiago, Chile 
	\and Instituto de Astrof\'{i}sica de Canarias (IAC), V\'{i}a L\'{a}ctea s/n, E-38206 La Laguna, Tenerife, Spain 
	\and Departamento de Astrof\'{i}sica, Universidad de La Laguna (ULL), E-38205 La Laguna, Tenerife, Spain
	}

   \date{Received 2013 November 8; accepted 6 February 2014}

 


 

  \abstract
  {Gravity modifies the spectral features of young brown dwarfs (BDs).  A proper characterization of these objects is crucial for the  identification of the least massive, and latest-type objects in star-forming regions, and to explain the origin(s) of the peculiar spectro-photometric properties of young directly imaged extrasolar planets and BD companions.}
   {We obtained medium-resolution (R$\sim$1500-1700) near-infrared (1.1-2.5 $\mu$m) spectra of seven  young M9.5-L3 dwarfs classified at optical wavelengths. We aim to empirically confirm   the low surface gravity of the objects in the near-infrared. We also  test whether self-consistent atmospheric models correctly represent the formation and the settling of dust clouds in the atmosphere of young late-M and L dwarfs.}
   {We  used ISAAC (Infrared Spectrometer And Array Camera) at VLT (Very Large Telescope) to obtain the spectra of the targets. We compared them  to those of mature and young BDs, and young late-type companions to nearby stars with known ages, in order to identify and study gravity-sensitive features. We computed  spectral indices weakly sensitive to the surface gravity to derive near-infrared  spectral types. Finally, we found the best fit between each spectrum and   synthetic spectra from the  BT-Settl 2010 and 2013 atmospheric models. Using the best fit, we derived the atmospheric parameters of the objects and identify which spectral characteristics the models do not reproduce.}
   {We confirmed that our objects are young BDs and we found near-infrared spectral types in agreement with the ones determined at optical wavelengths.  The spectrum of the $\mathrm{L2{\gamma}}$ dwarf \object{2MASSJ232252.99-615127.5} reproduces well the spectrum of the planetary mass companion \object{1RXS J160929.1-210524b}.  BT-Settl models fit the spectra and the 1-5 $\mu$m spectral energy distribution of the L0-L3 dwarfs for temperatures between 1600-2000 K. But the models fail to reproduce the shape of the H~band, and the near-infrared slope of some of our targets. This fact, and the best fit solutions found with super-solar metallicity are indicative of a lack of dust, in particular at high altitude,  in the cloud models.}
   {The modeling of the vertical mixing and of the grain growth will be revised in the next version of the BT-Settl models. These revisions may suppress the remaining non-reproducibilities.  Our spectra provide additional templates for the characterization of the numerous young L-type companions that will be detected in the coming years by planet imaging instruments such as VLT/SPHERE, Gemini/GPI, Subaru/SCexAO, and LBTI/LMIRCam.}


  \maketitle
%

\section{Introduction}\label{introduction}
Since the discovery of the first substellar objects (\citealt{1995Natur.378..463N,Rebolo1995}), large infrared (IR) surveys have unearthed hundreds of brown dwarfs (BDs) in the field and in star forming regions. The spectral energy distributions of BDs peak in the infrared. Their spectra are dominated by broad, overlapping condensate and molecular absorption features \citep[see][]{Kirkpatrick}.  The strength of these features depends on a combination of photospheric temperature, gas, pressure and dust properties, which in turn are related to the effective temperature ($\mathrm{T_{eff}}$), surface gravity ($g$), and metallicity of the objects. Spectroscopic studies of mature ($\gg 400$ Myr) field BDs led to the definition of the spectral classes `L' \citep{1999ApJ...519..802K, 1999AJ....118.2466M} for objects dominated by H$_{2}$O, FeH, and CO absorption bands in the near-infrared (NIR), and `T'  \citep{2002ApJ...564..421B, 2002ApJ...564..466G} for objects which exhibit strong CH$_{4}$+H$_{2}$O bands and collision-induced absorption (CIA) due to H$_{2}$ at these wavelengths. The change in the spectral morphology,  used for the classification, is mostly tied to changes in $\mathrm{T_{eff}}$ \citep[e.g.,][]{2009A&A...503..639T}.

Substellar objects contract and cool down with time \citep{1997ApJ...491..856B, BCAH98}. Therefore, young objects can have larger radii and higher luminosities than older and more massive objects despite an identical effective temperature. The lower surface gravity of young objects ($g \varpropto M/R^{2}$) can be directly accessed by  observation and be used to break the degeneracy.  Low surface gravity results in peculiar  spectral characteristics such as the triangular H band shape in the near-infrared \citep{Zapatero-Osorio2000, Lucas}, and reduced alkali lines in the optical and near-infrared \citep[e.g.,][]{McGovern, Cruz}.  These modifications have nonetheless mostly been investigated for objects earlier than $\sim$L4 found in young clusters and star forming regions \citep[e.g.,][]{Luhman1997, 1999ApJ...525..466L, Lucas,  2004ApJ...617..565L, 2004AJ....127..449M, Luhman2004, McGovern, Lodieu, Weights, 2012A&A...539A.151A}.  

Advances in  high contrast and high resolution imaging in the near-infrared (1-5 $\mu$m; NIR) led to the discovery of late-type companions to young nearby stars  straddling the planet/BD boundary \citep[e.g.,][]{2004A&A...425L..29C, 2005A&A...438L..29C, 2008ApJ...689L.153L, 2008Sci...322.1348M, 2010Natur.468.1080M, 2010Sci...329...57L, 2011ApJ...726..113I, 2013A&A...553L...5D, 2013ApJ...772L..15R, 2013ApJ...774...11K, 2013arXiv1310.4825C}. Most of these objects have estimated effective temperatures similar to those of L dwarfs. The low to medium-resolution (R$\leq$5000) spectra and infrared photometry of these objects exhibit peculiar features \citep[red pseudo-continua, triangular H~band shape, lack of methane absorption, reduced K I and Na I lines; ][]{2008ApJ...689L.153L, Bonnefoy2010, Patience, 2011ApJ...729..139W, Barman, Bonnefoy2013, 2013ApJ...768...24O}. These peculiarities are likely  directly related to the expected low surface gravity of the objects. 

Nevertheless, the  currently limited sample of young late-type objects with high S/N (signal-to-noise) spectra make the establishment of a proper empirical classification scheme challenging for these objects. \citet{Cruz} have identified a population of young bright and nearby  L dwarfs isolated in the field. They  developed a classification scheme in the optical. This scheme has been extended in the NIR for young L dwarfs by \citet{Bonnefoy2013} and \citet{Allers2013}. Several of these peculiar L-type dwarfs share similarities with the spectra of young companions at these wavelengths \citep{Bonnefoy2010, Bonnefoy2013, Allers2013, 2013AJ....145....2F, 2013arXiv1310.0457L}.  Getting more NIR spectra of young isolated objects are therefore needed to consolidate the classification scheme and to identify  further analogues to directly imaged exoplanets.

Atmospheric models  allow us to disentangle the effect of varying $T_{eff}$, log~\textit{g}, and (metallicity) \textit{M/H} on the spectral features. Below $T_{eff}$ $\sim$~2600~K, models predict that clouds of iron and silicate grains begin to form, changing the opacity (\citealt{Lunine}, \citealt{Tsuji}, \citealt{Burrows_Sharp}, \citealt{Lodders}, \citealt{Marley2000}, \citealt{Marley2001}, \citealt{Allard2001}). The formation and the gravitational sedimentation of these dust clouds are  influenced by the surface gravity. Dust cloud formation is  expected to be more efficient at low gravity because the atmosphere is more extended, and the gas cooler. Low gravity tends to make the convection and the resulting mixing more efficient as well.

Modeling of the spectro-photometric data on young L and early-T type companions with parametrized models (\citealt{Marley_Ackerman}, \citealt{Burrows}) has revealed anomalously thick clouds (\citealt{Barman2011a},  \citealt{Barman2011b}, \citealt{Skemer}, \citealt{Marley2013}). These peculiar cloud properties may explain why some companions are "underluminous" in some bands (\citealt{Skemer}, \citealt{Marley2012}). Self-consistent atmospheric models, such as the BT-Settl models \citep{Allard} and the Drift-PHOENIX models, \citep{Helling} use cloud models where the dust properties do not require defining any additional free parameters other than log \textit{g}, $T_{eff}$, \textit{M/H}. Synthetic spectra for a specific set of atmospheric parameters can be compared to empirical spectra. These models are just starting to be tested  on spectra of young late-type objects \citep[companions and free-floating objects;][]{Bonnefoy2010, 2011A&A...529A..44W, 2012A&A...540A..85P, Bonnefoy2013}. We used them on spectra of young M5.5-L0 dwarfs in \citet{Bonnefoy2013}  to reveal a drop of the effective temperature at the M/L transition.  We suggested that this drop could be induced by an improper handling of the formation of dust clouds at the M/L transition in the models. The test however could not be extended to later spectral types and lower effective temperatures due to the lack of a consistent sample of young objects in the L dwarf regime at that time. 

In this paper, we present a homogeneous set of seven medium-resolution (R$\sim$1500-1700) spectra of M9.5-L3 dwarfs, all classified at optical wavelengths. Our sample is composed of the M9.5 object \object{DENIS-P J124514.1-442907} (also called TWA 29; hereafter DENIS J1245) a member of TW-Hydrae (5-10~Myr), and  the L0 dwarf  \object{Cha J1305-7739} \citep[][hereafter Cha 1305]{Jayawardhana}, one of the least massive objects of the Chameleon II cluster. We also present the spectra of five L dwarfs with features indicative of low surface gravity (L$\gamma$ dwarfs) in the optical, identified by \cite{Cruz}. These objects are the two L0$\gamma$ dwarfs \object{EROS~J0032-4405} \citep[][hereafter EROS~J0032]{Goldman} and  \object{2MASS J22134491-2136079} \citep[][hereafter 2M2213]{Cruz}, the  L2$\gamma$ dwarf \object{2MASSJ232252.99-615127.5} \citep[][hereafter 2M2322]{Cruz}, and the two  L3$\gamma$ dwarfs \object{2MASS J212650.40-814029.3} \citep[][2M2126]{Cruz} and \object{2MASSJ220813.63+292121.5} \citep[][hereafter 2M2208]{Cruz}. 

 We aim to use the spectra to confirm the low surface gravities of the objects in the near-infrared and to test the ability of the BT-Settl models to correctly handle the formation and gravitational settling of dust under reduced surface gravity conditions.  These spectra enrich the scarce sample of empirical near-infrared medium-resolution spectra of young late-type objects  beyond the M-L transition, especially for spectral type L3.  We describe in Section \ref{observations} our observations and the associated data reduction. We present in Section \ref{empirical_analysis} an empirical analysis of the spectral features in order to derive  near-infrared spectral types and confirm the young age of our targets. In Section \ref{Spectralsynthesis} we describe the comparison of the atmospheric models to  the observed spectra. We discuss these comparisons  and derive updated target properties in Section \ref{discussion}.


\section{Observations and data reduction}\label{observations}
Our targets were  observed with the Infrared Spectrometer And Array Camera \citep[ISAAC, ][]{1998Msngr..94....7M} mounted on the VLT/UT3 telescope. The instrument was  operated  in low-resolution mode with the 0.3" slit at central wavelengths 1.25~$\mu$m, 1.65~$\mu$m and 2.2~$\mu$m. This setup provides spectra with resolving powers  of $\sim$1700, 1600, and 1500  from 1.1--1.4 $\mu$m (J band), 1.42--1.82 $\mu$m (H band), and 1.82--2.5 $\mu$m (K~band). Nodding and small jittering of the source along the slit were performed to correct bad pixels and to subtract the sky and bias contributions. Sources at high airmasses were observed with the slit aligned with the parallactic angle to mitigate differential flux losses.

We observed DENIS~1245  on April 6 and April 21, 2010. We took $6\times120$ s exposures in the J band, and $6\times90$ s exposures in the H and K bands.  We moved the star along the slit between two positions separated by 20 arcseconds following an ABBA strategy. We applied additional small offsets (5") around the two source positions between each exposure  to efficiently filter out non-linear and hot-pixels at the data reduction step. We followed a similar strategy for the remaining targets. Data integration times and the number of exposures are reported in Table~\ref{table_observations0}. Early-type stars were observed soon after the science target at similar airmass to ensure a proper removal of  telluric features.  These observations are also summarized in Table~\ref{table_observations0}. Calibrations were obtained during the day following the observations: flat fields, wavelength calibration frames, and frames with  a star moving along the slit in low and medium-resolution to compute the spectral curvature. \\

Data were reduced using the 6.1.3 version of the ISAAC pipeline \citep{1999Msngr..95....5D, 2004Msngr.118....2S} provided by the \textit{European Southern Observatory}. The pipeline identifies pairs of sky-object frames and subtracts them. The calibration in wavelength  and  the slit curvature distortion were performed using exposures with the Xenon and Argon lamps. The slit curvature was modeled with a bivariate 2-d polynomial. The dispersion relation was computed by matching a Xenon and/or Argon atlas with the corrected spectra. The pipeline divides the raw images by the flat field, corrects  bad-pixels and   distortion, and reconstructs the combined sky-subtracted 2D spectra from a shift of the nodded images. The object flux is extracted in each spectral channel to build the final spectrum. 

Data on the objects and associated telluric standard stars were reduced in a similar way. Telluric standard star spectra were divided by a black-body with a temperature that corresponded to their spectral type \citep{1991Ap&SS.183...91T}. He and H lines were interpolated in the resulting spectra using a low-order Legendre polynomial fit of the pseudo-continuum around the line. This produced the final estimate of the combined atmospheric and instrumental transmission. We obtained the final J, H, and K band spectra of the science targets by dividing them by this transmission. 

We created 1.1-2.5 $\mu$m flux-calibrated  spectra of the objects  using the following procedure. The  J, H, and K band ISAAC spectra were convolved with the filter transmission curves. The resulting spectra were integrated. We applied the same procedure to a  flux calibrated spectrum of Vega \citep{1985A&A...151..399M, 1985IAUS..111..225H}. We  then searched for the normalization factor of the ISAAC spectra that could produce a contrast ratio between the integrated flux of the science target and Vega which corresponds to  the J, H, and $K_{s}$-band photometry of the objects taken from 2MASS \citep[]{2003yCat.2246....0C, Allers2006}.

\begin{table*}[!h]
\begin{minipage}{18cm}
\caption{Observing log. $\lambda_{ref}$ is the central wavelength of the band, DIT is the integration time in each position of the slit and NDIT is the number of exposures. }  
\label{table_observations0}
  \centering
\begin{tabular}{llllllll}
 \hline
 \hline
Name & Date & $\lambda_{ref}$ & DIT & NINT & Seeing  & Airmass  & Notes \\
			&      &   ($\mu$m)   &  (s)  &   &   ('')  &   &   \\
 \hline              
DENISJ1245  	& April 4, 2010& 1.25 & 120 & 6 & 1.30  & 1.20   &  \\

DENISJ1245  	& April 21, 2010& 1.25 & 120 & 6 & 0.80 & 1.07    &  \\
DENISJ1245  	& April 21, 2010 & 1.65 & 90  & 6 & 0.70 & 1.06   &  \\ 
DENISJ1245  	& April 21, 2010 & 2.20 & 90  & 6 & 0.55 & 1.06   &   \\

HIP~064550        & April 4, 2010 & 1.25 & 5   & 2 & 1.15 & 1.20  &  G1.5V Telluric Standard \\

HIP~055667        & April 21, 2010& 1.25 & 5   & 2 & 0.70 & 1.07 &  B2IV-V Telluric Standard \\
HIP~055667        & April 21, 2010& 1.65 & 5   & 2 & 0.70 & 1.07 &  B2IV-V Telluric Standard \\
HIP~055667        & April 21, 2010& 2.20 & 5   & 2 & 0.60 & 1.07 &  B2IV-V Telluric Standard \\

\hline	
EROS~J0032   		& July 6, 2010 	& 1.25 & 180 & 6  & 1.30 & 1.07 & \\

EROS~J0032		& July 15, 2010 & 1.65 & 120 & 6 & 0.85 & 1.07 & \\
EROS~J0032		& July 15, 2010 & 2.20 & 120 & 6 & 0.85 & 1.07 & \\

EROS~J0032		& July 25, 2010 & 1.25 & 180 & 6  & 0.70 & 1.10 & \\

HIP~004722		& July 6, 2010  & 1.25 & 5    & 2 & - & 1.08 &  G3V Telluric Standard \\

HIP~111085      	& July 15, 2010 & 1.65 & 5    & 2 & 1.02 & 1.16 & B9V Telluric Standard \\
HIP~111085      	& July 15, 2010 & 2.20 & 5    & 2 & 1.04 & 1.16 &  B9V Telluric Standard \\

HIP~003356      	& July 25, 2010 & 1.25 & 5    & 2 & 0.80 & 1.07 &  B9.5V Telluric Standard \\

\hline
Cha 1305 		& April 6, 2010       & 1.65 & 200  & 6 & 0.90 & 1.90 & \\

Cha 1305 		& April 21, 2010      & 1.25 & 240  & 8 & 0.60 & 1.67 & \\
Cha 1305 		& April 21, 2010       & 1.65 & 200 & 6  & 0.50 & 1.66 & \\
Cha 1305 		& April 21, 2010      & 2.20 & 180 & 10 & 0.80 & 1.70 & \\

HIP~059830 	&   April 6, 2010        & 1.65 & 5   & 2 & 0.80 & 1.97&  B3V Telluric Standard \\

HIP~072671	&   April 21, 2010      & 1.25 & 5    & 2 &  -  & 1.60&  B8V Telluric Standard \\
HIP~072671	&  April 21, 2010      & 1.65 & 5    & 2 & 0.59 & 1.56&  B8V Telluric Standard \\
HIP~072671	&  April 21, 2010       & 2.20 & 5    & 2 & 0.75 & 1.56& B8V Telluric Standard \\

\hline
2M2322		&     June 9, 2010       & 1.25 & 240  & 6 & 1.00 & 1.35& \\
2M2322		&     June 9, 2010       & 1.65 & 120  & 10 & 0.80 & 1.28& \\
2M2322		&    June 9, 2010        & 2.20 & 120  & 10 & 1.00 & 1.26& \\

HIP~117661	&   June 9, 2010        & 1.25 & 5    & 2 & 0.70 & 1.40& B9V Telluric Standard \\
HIP~117661	&  June 9, 2010        & 1.65 & 5    & 2 & 0.90 & 1.39&  B9V Telluric Standard \\
HIP~117661	&  June 9, 2010        & 2.20 & 5    & 2 & 1.00 & 1.33&  B9V Telluric Standard \\

\hline
2M2126		&  April 21, 2010      & 1.65 & 180  & 6 & 0.60 & 2.02& \\
2M2126		& April 21, 2010       & 2.20 & 180  & 4 & 0.60 & 1.99& \\
2M2126		&  June 7, 2010      & 1.25 & 180  & 12 & 0.90 & 1.84& \\
HIP~112781	& April 21, 2010       & 1.65 & 5    & 2 & 1.10 & 2.08&  B6IV Telluric Standard \\
HIP~112781	& April 21, 2010       & 2.20 & 5    & 2 & 0.80 & 2.03&  B6IV Telluric Standard \\
HIP~099400	& June 7, 2010       & 1.25 & 5    & 2 & 0.80 & 1.73&  B2IV Telluric Standard \\

\hline
2M2208		& June 9, 2010         & 1.25 & 180  & 12 & 0.80 & 1.90& \\
2M2208		& June 7, 2010      & 1.65 & 180  & 6 & 0.80 & 1.70& \\
2M2208		& June 7, 2010       & 2.20 & 180  & 9 & 0.90 & 1.72& \\

HIP~112235		& June 9, 2010        & 1.25 & 5    & 2 & 0.70 & 1.97& B9V Telluric Standard \\
HIP~112235		&  June 7, 2010      & 1.65 & 5    & 2 & 0.95 & 1.73&  B9V Telluric Standard \\
HIP~112235		&  June 7, 2010      & 2.20 & 5    & 2 & 0.80 & 1.74& B9V Telluric Standard \\
\hline
2M2213   	   &   August 6, 2013   &   1.25  &   180  &  8  &  2.07   &  1.10   &   \\
2M2213   	   &   August 6, 2013   &   1.65  &   110  &  2  & 1.97   &  1.21 &   \\
2M2213   	   &   August 6, 2013   &   2.20  &   150  &  2   &   \dots  &  1.22 & \\

HIP~114656	   &   August 6, 2013   &   1.25  &  5  &  2  &  2.27   &  1.04   &   B9V Telluric Standard \\
HIP~114656	   &   August 6, 2013   & 1.65   &   5  &  2  & 1.71   &  1.05 &  B9V Telluric Standard \\
HIP~114656	   &   August 6, 2013   &  2.20   &   5  &  4   &  \dots   &  1.11 & B9V Telluric Standard \\

\hline

\end{tabular}
\end{minipage}
\end{table*}


\section{Empirical analysis}
\label{empirical_analysis}

In this section, we compare the spectral properties of our sample to  those of brown dwarfs and companions found in the literature in order to confirm features indicative of young age and assign near-infrared spectral types for the targets. We select the best fitting spectrum  using  $\chi^2$ minimization and  visual inspection over all wavelengths.

For that purpose, we used template spectra of young M and L-type companions (see the description in Appendix \ref{Appendix:A}),  late-M and early-L brown dwarfs from star forming regions and young nearby associations \citep[R from $\sim$120 to $\sim$11500;][]{2003ApJ...593.1074G, Slesnick, Allers2007, Lodieu, 2009ApJ...697..824A, Allers2010, Rice,Allers2013,  Bonnefoy2013,2013A&A...551A.107M}, young field L dwarfs \citep[R$\sim$1500;][]{Kirkpatrick2006, Allers2013, 2013arXiv1310.0457L}, and  MLT field dwarfs \citep[R$\sim$2000;][]{McLean, Cushing}. We also compared our spectra to low resolution templates (R$\sim$120) of the SpeX Prism Spectral Library\footnote{http://pono.ucsd.edu/$\sim$adam/browndwarfs/spexprism/}. The spectrum of Cha J1305 was dereddened by $A_{V}=3$ mag \citep{Allers2006} using the \cite{1999PASP..111...63F} extinction curve with an interstellar extinction parameter R(V)=3.1 prior to any comparison. We assumed $A_{v}=0$ for the remaining targets.
		
\subsection{An age-sequence of M9.5 dwarfs}  
\label{subsec:ageseq}

\begin{figure}
	\includegraphics[width=8cm]{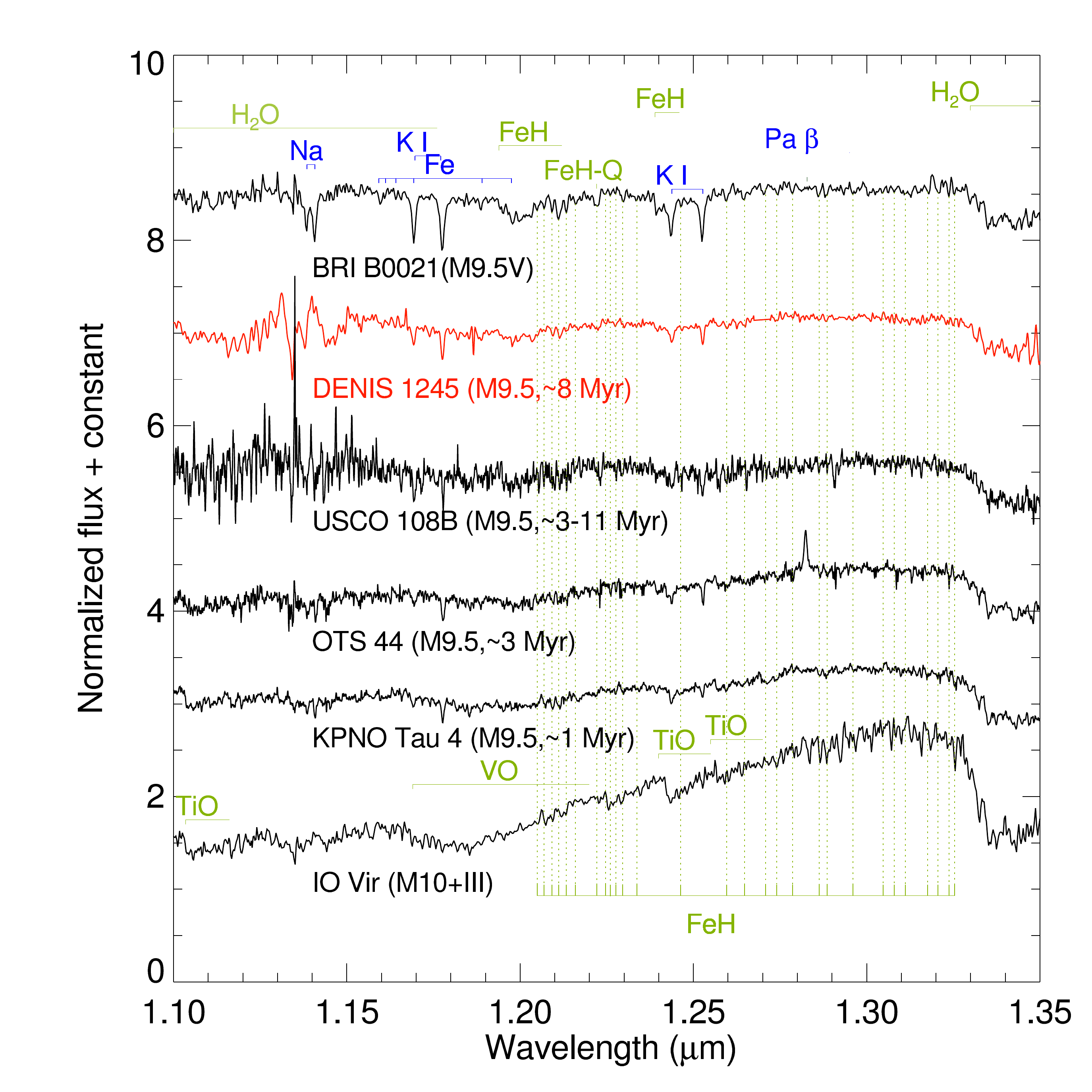}
	\caption{We plot spectra of six M9.5 objects with the same optical spectral type, but different ages, i.e gravity. We can appreciate the evolution of the spectral features with age in the J~band. The most remarkable ones are the appearance of alkali lines at older ages.}
\label{agesJ}
\end{figure}

\begin{figure}
	\includegraphics[width=7.9cm]{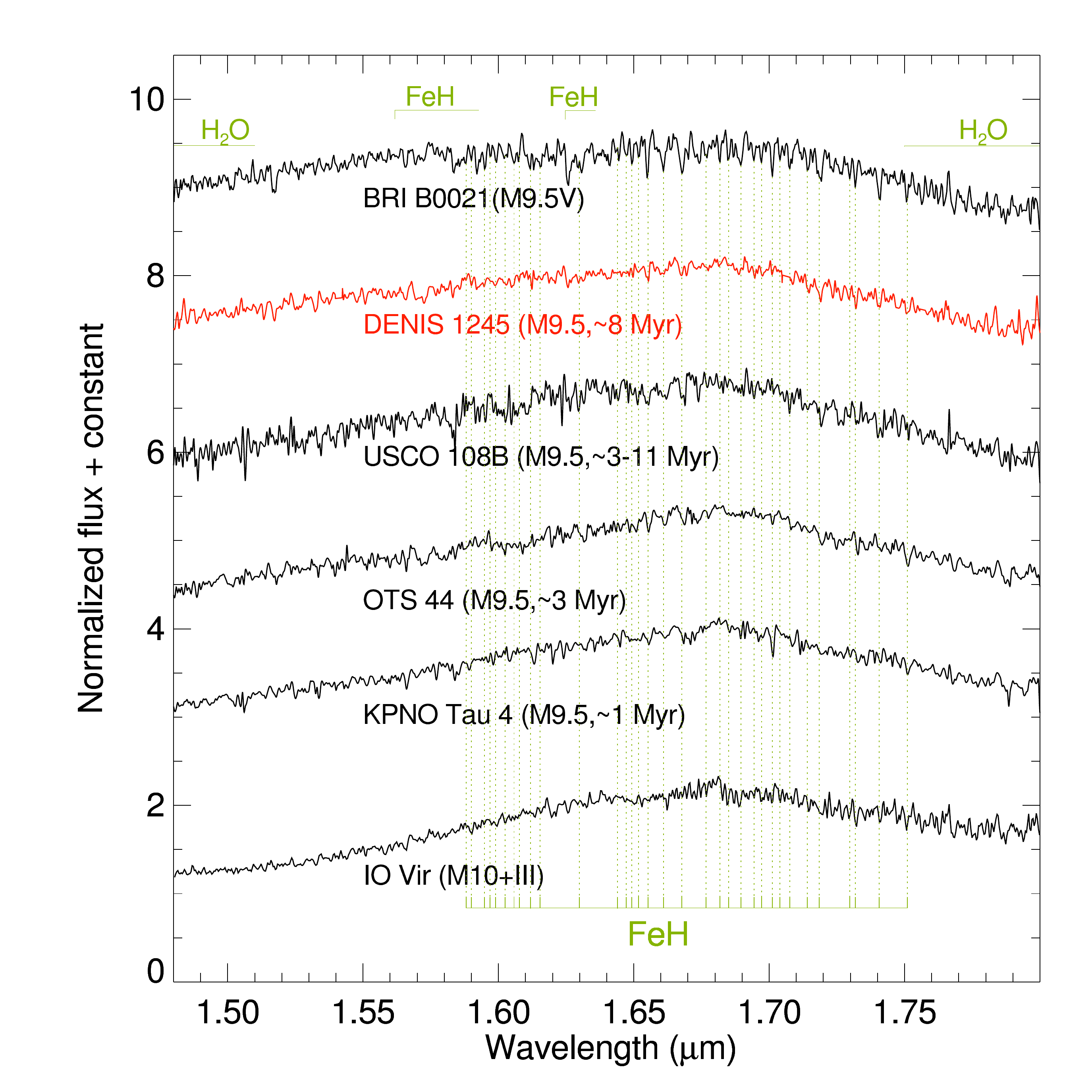}
	\caption{Same plot as Fig. \ref{agesJ}. Here we appreciate how the H~band becomes more triangular when we move to younger BDs.}
\label{agesH}
\end{figure}

\begin{figure}
	\includegraphics[width=8cm]{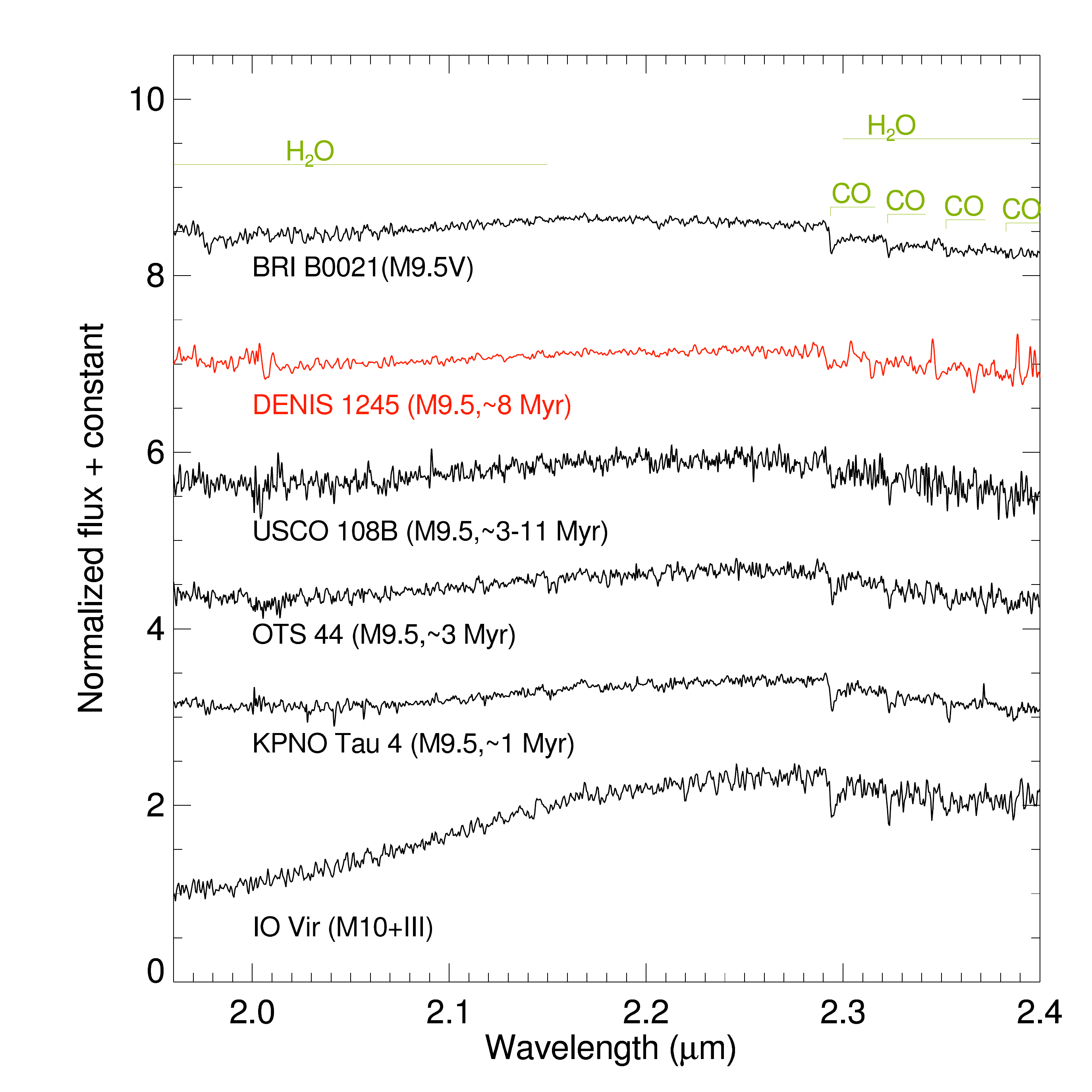}
	\caption{Similar plot to Fig. \ref{agesJ} and \ref{agesH} in the K band.}
\label{agesK}
\end{figure}

    The spectrum of DENIS-P~J124514.1-442907 complements an age-sequence of medium-resolution spectra of optically classified M9.5 dwarfs. We show in Figures \ref{agesJ}, \ref{agesH} and \ref{agesK} the J~band, H~band and K~band respectively with resolution of R$\sim$1500 in the H and K band and R$\sim$1700 in the J band. This resolution is sufficient to study atomic and molecular lines. The age-sequence is composed of spectra of members of the Taurus \citep[KPNO Tau 4; ][]{2002ApJ...580..317B}, Chameleon I  \citep[OTS 44; ][]{1999ApJ...526..336O, 2004ApJ...617..565L}, and Upper Scorpius  \citep[\object{UScoCTIO 108B}; ][]{2008ApJ...673L.185B} star forming regions. We also overlay the spectrum of the M10III Mira IO Virginis \citep{2009ApJS..185..289R} as an example of an extremely low surface gravity atmospheres. This sequence further highlights age-sensitive and gravity-sensitive features reported in the literature:  the increase of the alkali line depths (K I, Na I), an increase in the strength of the FeH absorptions at 1.20 and 1.24 $\mu$m, and of the $\mathrm{H_{2}O}$ band (1.7-2.25 $\mu$m).  The sequence also confirms the decrease of the VO band  strength from 1.17 to 1.22 $\mu$m and  that the J~band slope becomes bluer when the age increases, as shown by \cite{Kirkpatrick2006} (see Fig.~\ref{agesJ}). The other main changing feature is the progressive appearance of the triangular H~band profile when the age decreases, which corresponds to the water absorption profile (Fig.~\ref{agesH}). We used these characteristic features to confirm the young age and/or low surface gravity  of the L${\gamma}$ field dwarfs  in our sample.
   

\subsection{Young field L-dwarfs?}
\label{subsec:Lfield?}

\begin{figure}
	\resizebox{\hsize}{!}{\includegraphics{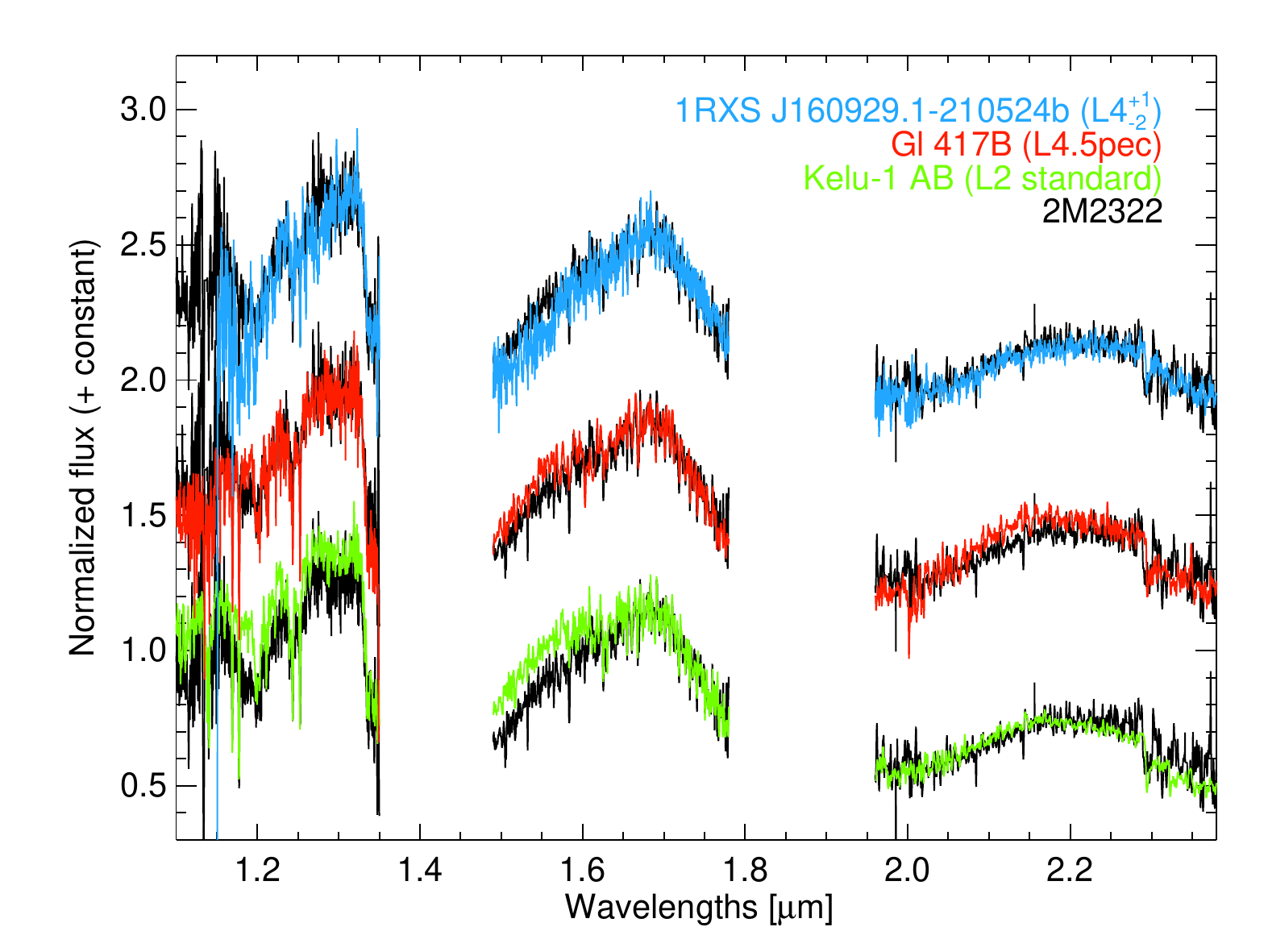}}
	\resizebox{\hsize}{!}{\includegraphics{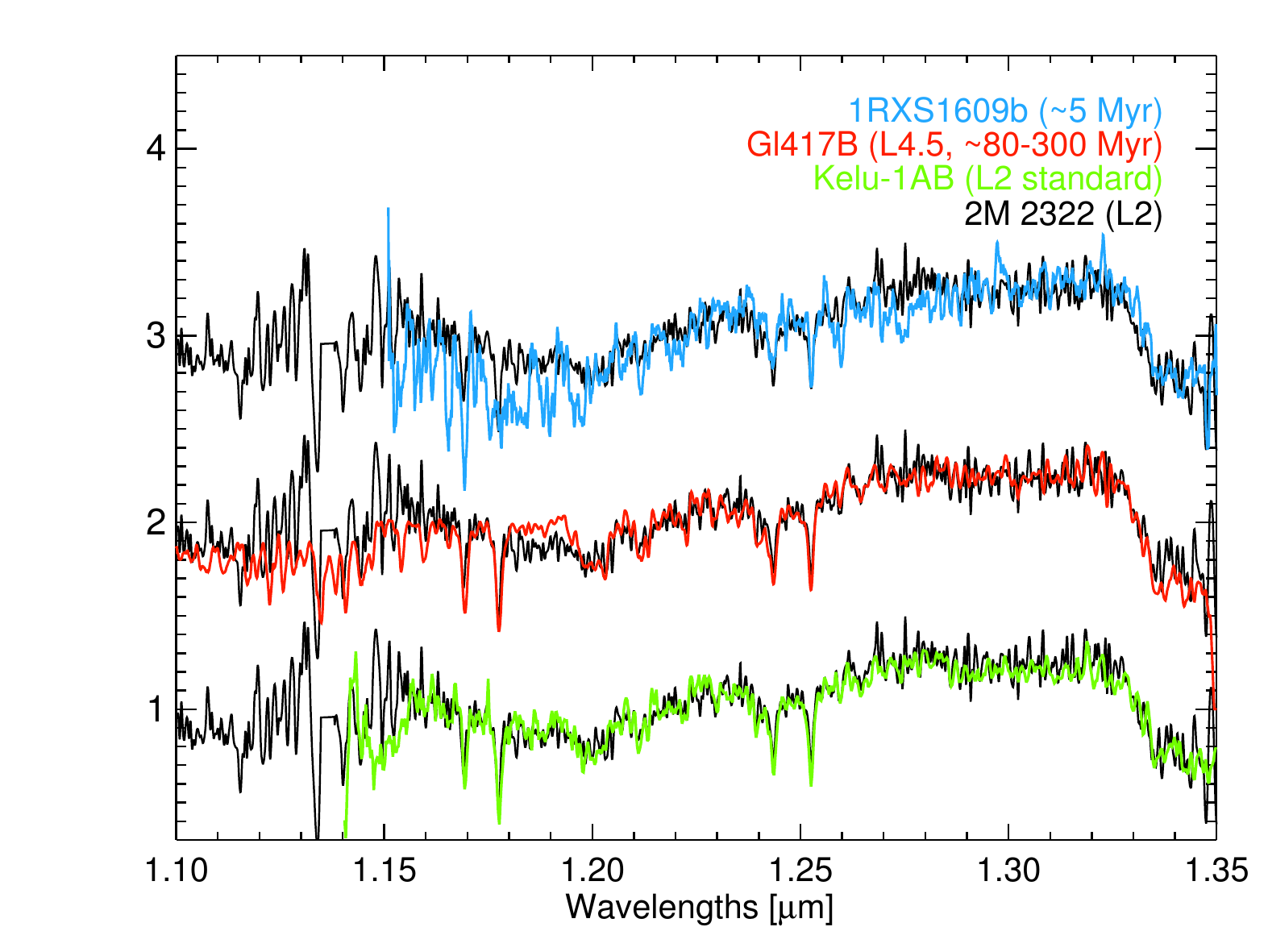}}
	\caption{In the top at the top we show the comparison of the 1.1-2.38 $\mu$m spectrum of the L2${\gamma}$ dwarf 2M2322 to the L2 optical standard \citep[Kelu-1 AB][]{1997ApJ...491L.107R, Cushing}, of the young  L4.5 binary companion GJ 417B \citep{2001AJ....121.3235K, Bonnefoy2013}, the planetary mass companion 1RXS J160929.1-210524b \citep{2008ApJ...689L.153L, 2010ApJ...719..497L}. In the plot at the bottom we show the J~band in a higher detail.}
\label{2M2322}
\end{figure}

In this section, we discuss the best fitting spectra for each L-dwarf of our sample. Among the sample of L dwarfs, the Chameleon member Cha~J1305-7739   shows clear features indicative of low surface gravity. This is in agreement with the age of the source. The remaining L dwarfs of our sample do not have assigned membership to young moving groups or clusters and clear ages. We therefore tried to identify features in the near-infrared typical of low surface gravity objects that would confirm the analysis derived by \cite{Cruz} in the optical. Plots showing the best matches  can be found in the Appendix \ref{Appendix:B} and the most remarkable result is shown in Figure \ref{2M2322}.

The near-infrared spectrum of the L0${\gamma}$ dwarf EROS~J0032 has spectral features midway between the medium-resolution (R$\sim$2000) IRTF (NASA Infrared Telescope Facility) spectrum \citep{Cushing} of the L1 field dwarf \object{2MASS J02081833+2542533} and of the M9.5 companion UScoCTIO 108B \citep{Bonnefoy2013}.  EROS~J0032 has a plateau from 1.59-1.69 $\mu$m characteristic of field dwarfs, but a more triangular shape in the H~band, and weaker FeH, and K I lines in the J band. Conversely, it has  deeper K I lines (1.169, 1.177, 1.243, 1.253 $\mu$m) and FeH bands at 1.2 $\mu$m and 1.624 $\mu$m than the companion. The 1.1-1.8 $\mu$m spectrum of the L0$\pm$1 companion to the Upper Scorpius star \object{GSC 06214-00210}  \citep{2011ApJ...726..113I, 2011ApJ...743..148B} reproduces perfectly the pseudo-continuum shape of EROS~J0032. The companion has nonetheless slightly weaker FeH and K I absorptions. The near-infrared spectral-slope of EROS~J0032 is redder than the one of the L0 field dwarf standard \object{2MASS J03454316+2540233} \citep{McLean, Kirkpatrick2010}. This can be attributed to reduced CIA of H$_{2}$ \citep{1997A&A...324..185B, Kirkpatrick2006}, and therefore low surface gravity.  EROS~J0032 also has reduced FeH lines at 1.2 $\mu$m and a more triangular H~band shape than the standard. Nevertheless, the two spectra have comparable K I line depths. The SINFONI (Spectrograph for INtegral Field Observations in the near-infrared) spectrum of benchmark  L0${\gamma}$ \object{2MASS J01415823-4633574} \citep{Bonnefoy2013} has a more triangular H~band shape, weaker FeH absorptions, and K I lines than the object. \cite{Allers2013} and \cite{2013arXiv1309.6525M} also found indication for low surface gravity in near-infrared spectra of EROS~J0032.  The lower resolution spectrum (R$\sim$100) of EROS~J0032 of \cite{Allers2013} matches perfectly our ISAAC spectrum. \cite{Allers2013} assign the same gravity class for this target as for AB Pic b, a $\sim$30 Myr old low-mass L0 \citep{Bonnefoy2013} companion.  Here, we conclude that EROS~J0032 only shows \textit{moderate} signs of  low surface gravity. 

EROS~J0032 and the second L0${\gamma}$ dwarf of our sample 2M2213 have comparable near-infrared spectral slopes. Nevertheless, we find that 2M2213 has a H~band shape, K I and FeH line depths midway between those of EROS~J0032 and UScoCTIO 108B. Therefore, this comparison suggests that 2M2213 has a lower surface gravity and a younger age than EROS~J0032, but is older than the assumed age of Upper Scorpius  \citep[$\sim$5-11~Myr;][]{Preibisch_Zinnecker, Slesnick2008, 2012ApJ...746..154P}. \cite{2013arXiv1309.6525M} also obtained a medium-resolution near-infrared spectrum of this object, and reach similar conclusions. They classify it as a L2pec based on the good match with the near-infrared spectrum of a field L2 standard. Nevertheless, such a comparison can not be done for young objects \citep[][]{2004ApJ...617..565L}, as it can lead  to later spectral type estimates. To conclude, we find that our spectrum perfectly matches the low resolution (R$\sim$100) spectrum of 2M2213 obtained by \cite{Allers2013}. 

In Figure \ref{2M2322} we show that the spectrum  of the L2${\gamma}$ dwarf 2M2322 is reproduced by  the spectrum \citep{Bonnefoy2013}  of the moderately old L4.5+L4.5 binary companion GJ417~B \citep[age from 80 to 890 Myr,][]{2001AJ....121.3235K, Allers2010}. But the object exhibits weaker FeH absorption, and K I lines than the companion. The  pseudo-continuum shape of 2M2322 from 1.95 to 2.5~$\mu$m, and from 1.45 to 1.6 $\mu$m  
is midway between the one of  GJ 417 B and of the 5-11 Myr old L4 planetary mass companion \object{1RXS J160929.1-210524b} \citep{2008ApJ...689L.153L, 2010ApJ...719..497L}. All three objects have similar near-infrared spectral slopes. 2M2322 also has a slightly redder spectral slope, more triangular H~band, and bumpy pseudo-continuum in the K band  
than the  L2 field dwarf standard \object{Kelu-1 AB} \citep{Cushing, Kirkpatrick2010}. All these comparisons provide evidence that this object has a low surface gravity. As a by-product, the companion suggests that 1RXS J160929.1-210524b has a spectral type L2. 

The 1.1-2.5 $\mu$m pseudo-continuum of 2M2126 is well reproduced by the spectrum of the  young \cite[$\sim$20-300~Myr][]{Zapatero-Osorio} $\mathrm{L3{\gamma}}$ companion  G~196-3B \citep[][]{Rebolo} gathered by \cite{Allers2013}. The depth of alkali-lines and of the FeH absorption at 1.2 $\mu$m are similar for the two objects. Nevertheless,  the water-band absorptions from 1.33-1.35 $\mu$m and 1.45-1.6 $\mu$m are deeper  in the companion spectrum.  The spectral slope from 1.2-1.33 $\mu$m is also redder in the spectrum of  G~196-3B. We also find a good match with the spectrum of the brown dwarf companion CD-35 2722B  \citep{2011ApJ...729..139W}, classified as L3 by \cite{Allers2013}. CD-35 2722 B forms a coeval system with CD-35 2722 A, a member of the 75-150 Myr old AB Doradus association.  We then conclude that 2M2126 is a young object, with indications of reduced surface gravity, and features in the near-infrared consistent with its optical class L$3{\gamma}$ determined by \cite{Cruz}.  

The second  L$3{\gamma}$ dwarf of the sample, 2M2208, has an identical spectrum to 2M2126. Both objects  show comparable K I and FeH line depths in the J band, and a similar gravity-sensitive slope from 1.45 to 2.29 $\mu$m. Nevertheless, this object has a bluer 1.1-2.5 $\mu$m slope than 2M2126 (and therefore G196-3B). The blue slope likely arises from an improper scaling of the J, H, K~band spectra due to uncertainty in the photometry. Indeed, the 1.1-1.35 $\mu$m portion of the spectrum has a slightly higher flux than the low-resolution (R$\approx$100) spectrum of the source  obtained by \cite{Allers2013}, but a similar pseudo-continuum otherwise. The slope is better reproduced by the spectrum  of GJ417 B   \citep{2001AJ....121.3235K, Bonnefoy2013} and by an IRTF spectrum of the L3 field dwarf \object{2MASS J15065441+1321060} \citep{2000AJ....119..369R, Cushing}. But the KI and FeH lines are weaker in the spectrum of 2M2208 than in the spectra of these two objects.  Therefore, the analysis confirms that 2M2208 shows signatures characteristics of young L3-L5 objects, and a spectral type similar to the one of 2M2126.

\subsection{Indices and equivalent widths}
\label{subsec:Indices}
To further assess the age, surface gravity, and spectral classes of our targets, we  computed spectral indices and equivalent widths that  quantify the evolution of the main absorption features. 

\begin{figure}[top!]
	\resizebox{\hsize}{!}{\includegraphics{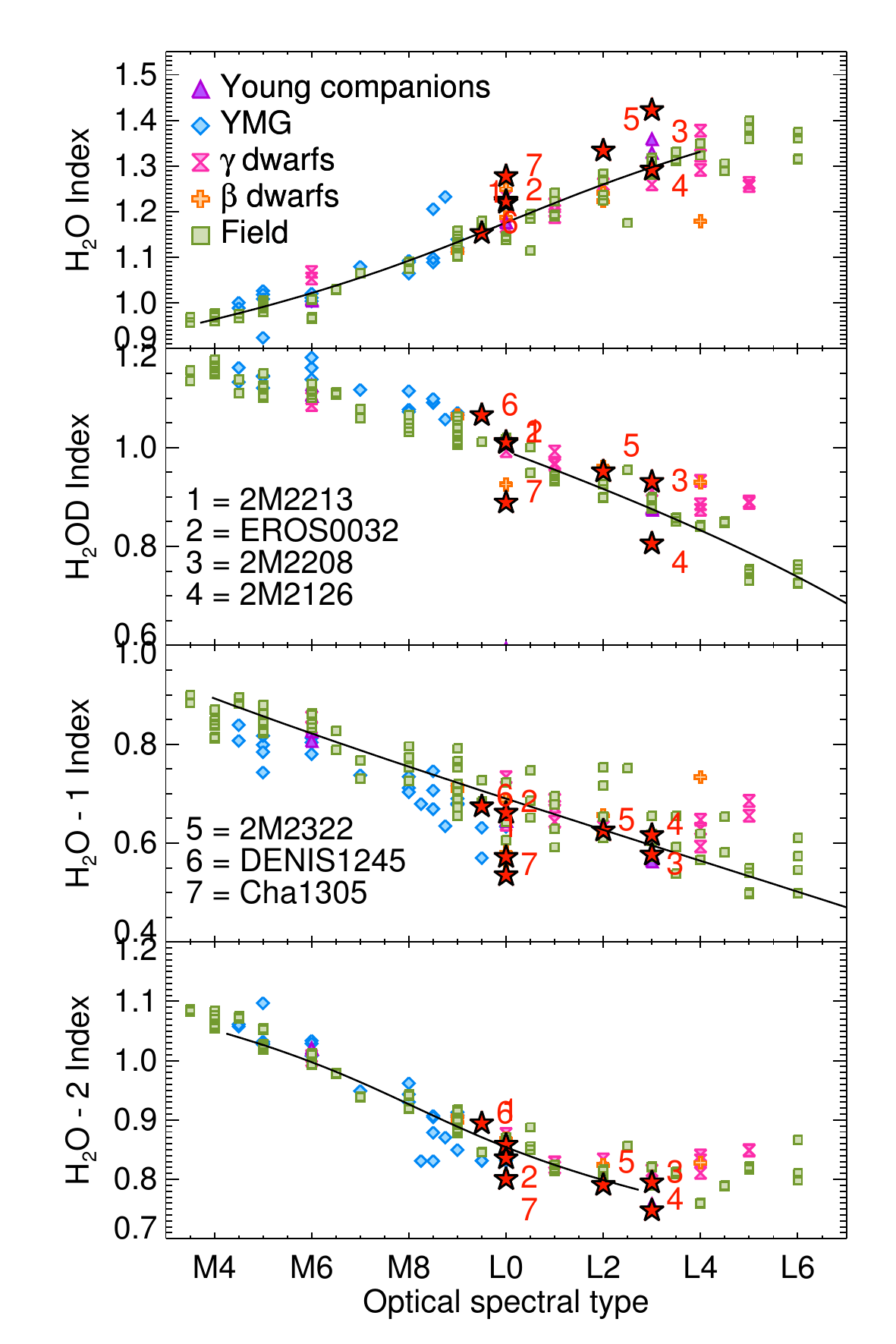}}
	\caption{Spectral indices used for the spectral type estimation of the NIR spectral type. Our targets are represented by red stars and are numerated from 1 to 6: 1 $=$ 2M2213, 2 $=$ EROS~J0032, 3 $=$ 2M2208, 4 $=$ 2M2126, 5 $=$ 2M2322, 6 $=$ DENIS~1245 and 7 $=$ Cha1305. The black lines represent the three degree polynomial fit to the field brown dwarfs. We report the index values for objects classified in the optical that are members of young clusters, star forming regions, or nearby associations (YMG), and for dwarfs with low ($\gamma$) and intermediate ($\beta$) gravity classes \citep{Kirkpatrick}.}
\label{Sptype_sensitive}
\end{figure}

We  first used spectral indices measuring the strength of the main water bands. These indices were selected independently by \cite{Bonnefoy2013} and/or \cite{Allers2013} from \cite{Allers2007} -- $H_{2}O$, \cite{Slesnick} --$H_{2}O$-1 and $H_{2}O$-2, and  \cite{McLean} -- $H_{2}OD$. They are known to show a clear trend with spectral type, and to be only weakly sensitive to the age or to the gravity class ($\gamma$, $\beta$). We computed them on  the compilation of  near-infrared spectra of young M3-M9.5 dwarf members of star forming regions (1-11 Myr) and young nearby associations (age $<$50 Myr)  classified in the optical.  We also derived them for young L${\gamma}$, L${\beta}$ dwarfs, and companion spectra provided by \cite{Allers2013}  and \cite{Bonnefoy2013}, and for field dwarfs obtained by \cite{McLean} and \cite{Cushing}. We degrade the resolution of all the spectra to R$\sim$100, which is our lowest resolution. Results are shown in Figure \ref{Sptype_sensitive}. We re-ajusted a third-order polynomial function on these trends (Table~\ref{coefficients}), as in \cite{Allers2013}, and use them to derive spectral types estimates  (Table~\ref{Tab:Sptypes}). We calculated the errors in the spectral type  as the root mean square (rms). We estimated the final near-infrared spectral types taking the mean of the different estimates from the indices  weighted by the associated error, and their errors as the standard desviation. These estimates are all consistent with the optical spectral types. They are also consistent with the matches found in Section \ref{subsec:Lfield?}.

\begin{table*}[h!]
  \caption{Coefficient of the polynomial fit derived from index values computed on field dwarf spectra and shown in Figure \ref{Sptype_sensitive}}
  \label{coefficients}
\begin{center}
\begin{tabular}{l llllll l}
\hline
\hline
Index definition & References & SpT &\multicolumn{4}{c}{Coefficients of Polynomial Fits} & RMS Sp. type\\  
    &   &  & a &  b & c & d \\
 \hline
$H_{2}O=\frac{<F\_\lambda=[ 1.550,1.560]>}{<F\_\lambda=[ 1.492,1.502]>}$&\citet{Allers2007}& M4-L4 & -199.72 & 487.93 &-394.58 & 111.66&  1.0 \\
$H_{2}OD=\frac{<F\_\lambda= [1.951,1.977]>}{<F\_\lambda= [2.062,2.088]>}$&\citet{McLean}& L0-L6 & 20.16 & 10.28 &  -24.13 & 3.48 &  1.0 \\
$H_{2}O-1=\frac{<F\_\lambda= [1.335,1.345]>}{<F\_\lambda= [1.295,1.305]>}$ & \citet{Slesnick} & M4-L6 &29.42 & -17.94 & -24.98 & 14.73  &  0.5  \\
$H_{2}O-2=\frac{<F\_\lambda= [2.035,2.045]>}{<F\_\lambda= [2.145,2.155]>}$ & \citet{Slesnick} & M4-L3 & 306.25 & -917.19 &965.74 & -348.88  &  1.0 \\

\hline
\end{tabular}
\end{center}
\end{table*}



\begin{table*}
  \caption{Estimation of the near-infrared spectral types based on spectral indices}
  \label{Tab:Sptypes}
\centering
\renewcommand{\footnoterule}{}  
\begin{center}
\begin{tabular}{l l l l l l l l l l}
 \hline
 \hline
Name & Opt SpT & Reference & Empirical SpT &\multicolumn{4}{c}{Index SpT}  & NIR SpT \\
     &         &           &               & $H_{2}O$ & $H_{2}OD$ & $H_{2}O-1$ & $H_{2}O-2$   & \\
 \hline

DE~J1245  & M9.5 & 1    & M9                        & M9.5$\pm$1.0 & M8.0$\pm$1.0 & L0.5$\pm$0.5 & M9.0$\pm$1.0 & M9.5$\pm${1.0} \\

EROS~J0032  & {L0$\gamma$}   & 2,6  & L1     & L1.0$\pm$1.0 & L0.0$\pm$1.0 & L1.0$\pm$0.5 & L0.5$\pm$1.0 & L0.5$\pm${0.5}\\

2M~J2213     &  { L0$\gamma$}   &   8  & L0  &  L1.0$\pm$1.0 & M9.5$\pm$1.0 & L3.5$\pm$0.5 & L0.0$\pm$1.0 &L2.0$\pm${1.5} \\

Cha~J1305 & L0   &  3,7 & L1                        & L2.5$\pm$1.0 & L2.5$\pm$1.0 & L5.0$\pm$0.5 & L2.0$\pm$1.0 & L3.5$\pm${1.5}\\

2M~J2322  & {L2$\gamma$}   & 4, 8 & L2       & L4.0$\pm$1.0 & L1.0$\pm$1.0 & L2.0$\pm$0.5 & L2.5$\pm$1.0 & L2.0$\pm${1.0}\\

2M~J2126  & {L3$\gamma$}   & 4, 8 & L3       & L3.0$\pm$1.0 & L5.5$\pm$1.0 & L2.0$\pm$0.5 & L4.5$\pm$1.0 & L3.0$\pm${1.5}\\

2M~J2208  & {L3$\gamma$}   & 5, 8 & L1       & L7.0$\pm$1.0  & L2.0$\pm$1.0 & L3.5$\pm$0.5 & L2.0$\pm$1.0 & L3.0$\pm${2.0}\\

\hline
\end{tabular}
\end{center}
\tablefoot{[1] - \citet{Looper}, [2] - \citet{Reid}, [3] - \citet{Allers2006}, [4] - \citet{Reid}, [5] - \citet{Allen}, [6] - \citet{Allers&Liu}, [7] - \citet{Alcala}, [8] - \citet{Cruz}}
\end{table*}


\begin{table*}
 \begin{minipage}{\linewidth}
 \caption{Age-sensitive indices and equivalent widths}
  \label{EW_table}
\centering
\renewcommand{\footnoterule}{}  
\begin{tabular}{l l lll l l l}
\hline
\hline
Object & $\mathrm{FeH_{J}}$\tablefootmark{a}	  &  $\mathrm{K I_{J}}$\tablefootmark{a}		 & $\mathrm{H-cont}$\tablefootmark{a}		  &   EW  - K I & EW - K I  & EW - K I  & EW - K I  \\  
           &                                 &                               &                   &      1.169 $\mu$m    & 1.177 $\mu$m   & 1.243 $\mu$m   &   1.253 $\mu$m    \\  
           &                                 &                               &                   &          ($\AA$)           &   ($\AA$)             &  ($\AA$)             &    ($\AA$)               \\
 \hline
DE~J1245   & $1.091\pm0.005$ & $1.058\pm0.002$  & $0.992\pm0.001$  &  $2.20\pm0.57$ & $2.57\pm0.76$ & $2.05\pm0.13$ & $1.02\pm0.12$   \\
EROS~J0032 & $1.200\pm0.003$ & $1.096\pm0.005$  & $0.925\pm0.001$  &  $5.92\pm0.31$ & $9.30\pm0.41$ & $5.97\pm0.17$ & $4.46\pm0.16$ \\
2M~J2213   & $1.134\pm0.004$ & $1.034\pm0.004$  & $0.958\pm0.002$  &  $7.45\pm0.92$ & $12.45\pm1.31$& $1.31\pm0.28$ & $1.94\pm0.25$  \\
2M~J2322   & $1.144\pm0.009$ & $1.112\pm0.003$  & $0.935\pm0.001$  &  $3.86\pm0.56$ & $7.78\pm0.74$ & $5.78\pm0.21$ & $4.51\pm0.19$  \\
2M~J2126   & $1.183\pm0.010$ & $1.061\pm0.003$  & $0.949\pm0.001$  &  $4.98\pm0.34$ & $6.55\pm0.46$ & $4.43\pm0.22$ & $4.52\pm0.20$ \\
2M~J2208   & $1.118\pm0.010$ & $1.063\pm0.003$  & $0.930\pm0.001$  &  $7.69\pm0.41$ & $8.39\pm0.55$ & $7.43\pm0.39$ & $3.86\pm0.36$ \\
\hline
\end{tabular}
\end{minipage}
\tablefoot{ \tablefoottext{a}{Computed using the code developed by \cite{Allers2013}.} }

\end{table*}

\begin{figure}

	\resizebox{\hsize}{!}{\includegraphics{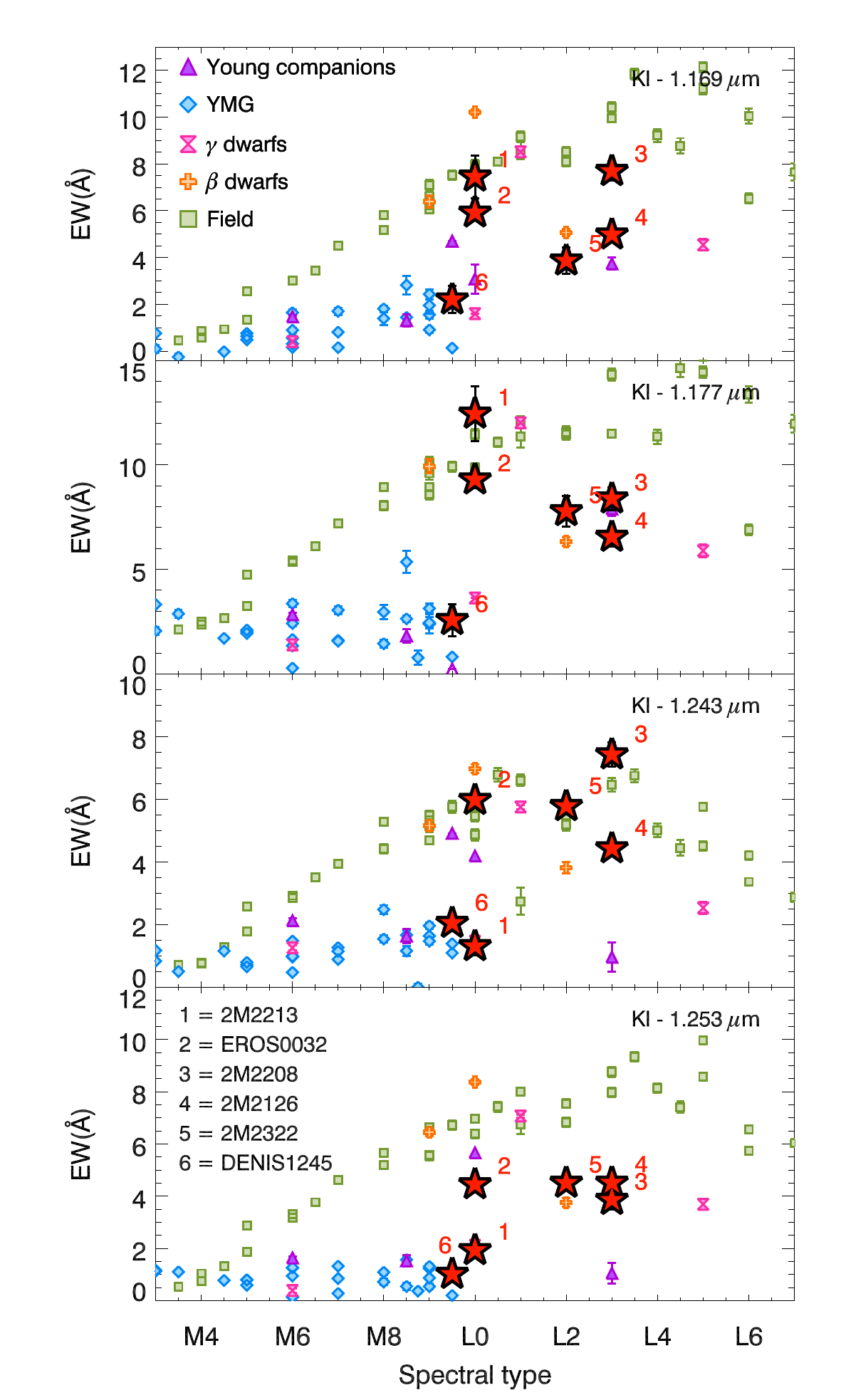}}
	\caption{Equivalent widths for the K I lines at 1.169~$\mu$m, 1.177~$\mu$m, 1.243~$\mu$m and 1.253~$\mu$m for our targets (red stars) and reference objects.}
\label{EW_KK}
\end{figure}

\begin{figure}
	\resizebox{\hsize}{!}{\includegraphics{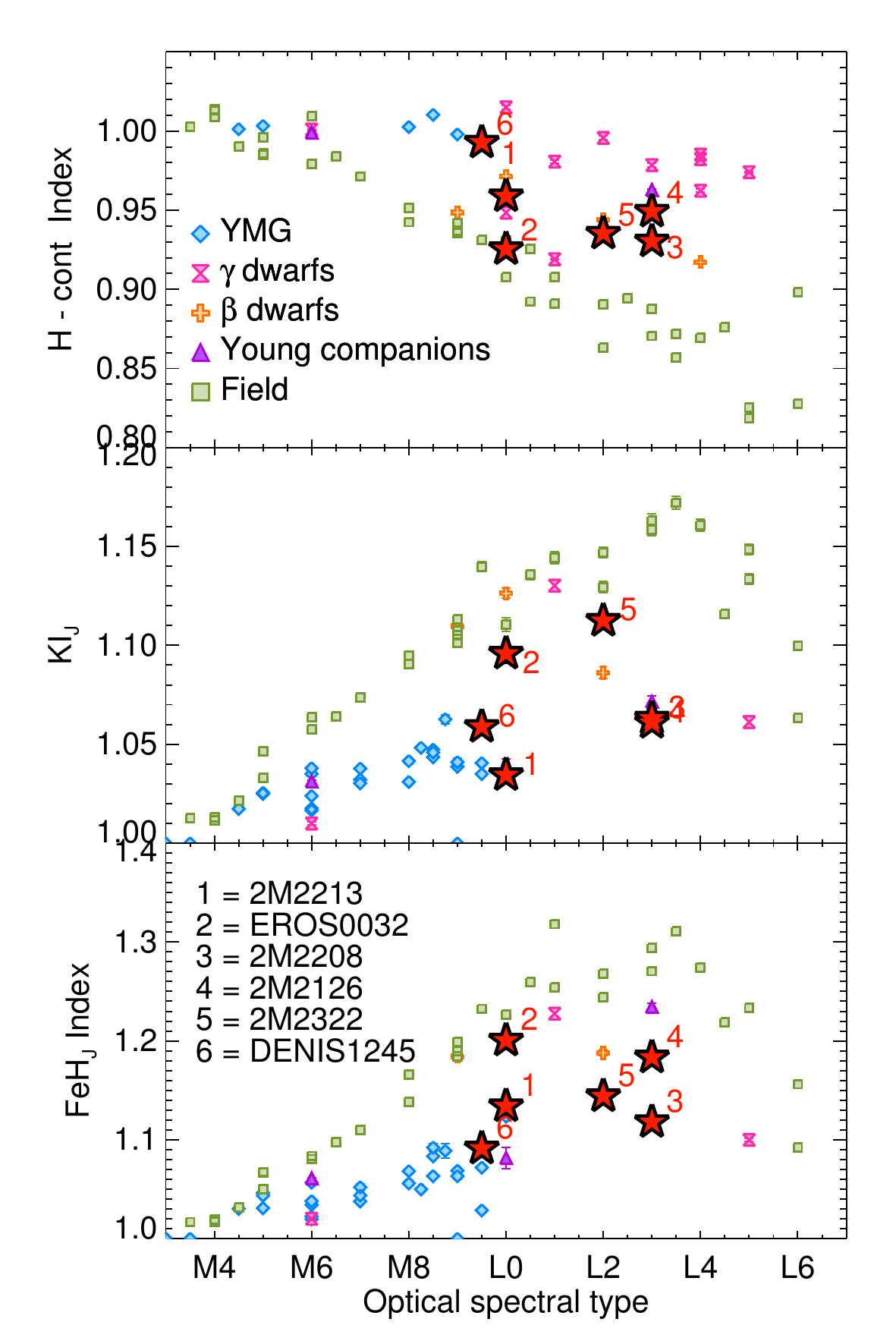}}
	\caption{FeH, KI$_{J}$, and H~band age-sensitive indices.}
\label{Gravity_sensitive1}
\end{figure}

The equivalent widths of the gravity-sensitive K I lines at 1.169~$\mu$m, 1.177~$\mu$m, 1.243~$\mu$m and 1.253~$\mu$m of our objects are reported in Table~\ref{EW_table}. They were computed following the method developed by \cite{1992ApJS...83..147S}. We  used the same reference wavelengths for the fit of the pseudo-continuum and for the line as \cite{Allers2013}. In Fig.~\ref{EW_KK} we show the equivalent widths of these lines for our targets and reference objects. The trends are similar to those found by \cite{Bonnefoy2013} and \cite{Allers2013}.  For all  four cases, EROS~J0032 has equivalent widths close to those of field L0 dwarf analogues. The remaining field L${\gamma}$ dwarfs have lower equivalent widths in some, but not all of the diagrams. This confirms the conclusions derived in Section \ref{subsec:Lfield?}. DENISJ1245 has  equivalent widths comparable to late-M dwarf members of the  $\sim$8 Myr old TW Hydrae, and Upper Scorpius. This is consistent with the membership of this object to the TW Hydrae association \citep[see][and Section \ref{DENIS_J1245}]{Looper}.  We calculate the $KI_{J}$ index for medium-resolution spectra degraded at R$\sim$700. We find similar results and trends (Figure \ref{Gravity_sensitive1}) with the $KI_{J}$ index defined by \cite{Allers2013}, and measuring the depth of the K I doublet at 1.243-1.253 $\mu$m. The spectrum of Cha~J1305 was too noisy to derive  equivalent widths and  $KI_{J}$ index values.

We also computed the $FeH_{J}$  and $H-cont$ indices defined by \cite{Allers2013}. These indices measure the strength of the gravity sensitive FeH feature at 1.2 $\mu$m and the shape of the H~band continuum, respectively. We calculate the  $FeH_{J}$ index using medium-resolution spectra degraded at R$\sim$700. Nonetheless, we use all the spectra smoothed to R$\sim$100 to calculate the $H-cont$ index, as the H~band is broad enough not to be  significantly affected by the spectral resolution. The index values are reported for the targets in Table~\ref{EW_table} and compared to other objects in Fig. \ref{Gravity_sensitive1}. All the objects, except EROS~J0032 have indices  compatible with the trends of objects from young moving groups and $\gamma$ and $\beta$ dwarfs. This further indicates that EROS~J0032 is the object with the highest surface gravity and/or the oldest age of the sample.

\section{Spectral synthesis}
\label{Spectralsynthesis}
  In the following sections, we compare the dereddened near-infrared spectra of our objects to predictions from  BT-Settl atmospheric models \citep{2003IAUS..211..325A, 2007A&A...474L..21A, 2011ASPC..448...91A}.   The models have already been tested  on near-infrared spectra of young M5.5-L0 objects by \cite{Bonnefoy2013}. We compare the new ISAAC spectra to derive the atmospheric parameters ($\mathrm{T_{eff}}$, log~g) of the objects and to reveal non-reproducibilities of the models for later spectral types. Flux in synthetic spectra  are provided per  $cm^{2}$ of the stellar surface. They therefore need to the scaled back, using the distance modulus, to absolute flux. 
The models are  described in \cite{2011ASPC..448...91A}, \cite{Allard2012a}, and \cite{Allard2012b}.  We summarize below their most relevant characteristics.
  
  The BT-Settl models account for the formation and gravitational settling of dust grains for $\mathrm{T_{eff}}$ below $\approx$2700 K in the photosphere of the objects following the approach described in \cite{1978Icar...36....1R}. The timescales of the main processes (mixing, sedimentation, condensation, coalescence and coagulation) are compared to determine the density distribution of grains and the  average grain size from the innermost to the outermost layers of the cloud.  One hundred and eighty types of condensates are accounted for in the models by their interaction with the gas phase chemistry, depleting the gas from their vapor phase counterparts. Fifty-five of these grain species are
included in the radiative transfer to the extent to which they have not settled from the cloud layer.

 The cloud model is implemented in the \texttt{PHOENIX} multi-purpose atmosphere code version 15.5 \citep{Allard2001}, which is used to compute the model atmospheres and generate synthetic spectra. Convective energy transport and velocities are calculated using Mixing Length Theory with a mixing length of 2.0 pressure scale heights, and overshoot is treated as an exponential velocity
field with a scale height based on the RHD simulations of  \cite{Ludwig2002,Ludwig2006} and \cite{Freytag2010,Freytag2012}; an additional advective mixing term due to gravity waves is included as described in \citet{Freytag2010}.  All relevant molecular absorbers are treated with line-by-line opacities in direct opacity sampling as in \citet{Allard2003}; relative to this the molecular line lists have been updated as follows: water-vapor \citep[BT2,][]{2006MNRAS.368.1087B}, CIA of $\mathrm{H_{2}}$ \citep{2002A&A...390..779B}, and FeH \citep{1998A&A...337..495P, 2003ApJ...582.1059W, 2003ApJ...594..651D}. Non-equilibrium chemistry for CO, $\mathrm{CH_{4}}$, $\mathrm{CO_{2}}$, $\mathrm{N_{2}}$ and $\mathrm{NH_{3}}$ is treated with height-dependent diffusion also based on the RHD simulation results of \cite{2010A&A...513A..19F}.

  
 We used the 2010 and 2013 pre-release of these models\footnote{The 2013 pre-release of the models corresponds to the final and stable version of the 2012 model grid (called BT-Settl 2012; used in \cite{2013A&A...555A.107B, Bonnefoy2013, 2013arXiv1308.3859B}) where some synthetic spectra were re-computed following the discovery of errors in the model code. Upgrades in the 2012 grid were released progressively on the \textit{Star, Brown Dwarf \& Planet Simulator} web server (\texttt{http://phoenix.ens-lyon.fr/Grids/BT-Settl/CIFIST2011
 /RESTARTS/}) until September 2013. Therefore, we decided to call the most recent version of the models used in this study \textit{BT-Settl 2013}, even if a new version of the BT-Settl models may be released on the web server by the end of 2013.}. The comparative analysis of the results provided by these two versions enables us to judge  the pertinence of new physics incorporated. This also avoids relying only on the most recent version of the models, which remain to be tested, contrary to the BT-Settl 2010 models \citep[e.g. ][]{Bonnefoy2013}.  In the 2013 pre-release of the BT-Settl models \citep[][hereafter BT-Settl 2013]{2011ASPC..448...91A}, the cloud model was improved by a dynamical determination of the supersaturation, the implementation of a grain size-dependent forward scattering, and by accounting for grain nucleation based on cosmic rays studies \citep{2005JASTP..67.1544T}. The BT-Settl 2010 models rely on the reference solar abundances of \cite{Asplund2009}. Conversely, the BT-Settl 2013 models are based on the \texttt{CIFIST} photospheric solar abundances of \cite{2011SoPh..268..255C}. Therefore, the 2013 solar-metallicity models have atmospheres slightly enriched with C, O compared to the 2010 models, e.g. in elements involved in the formation of the main molecular absorbers in the near-infrared (CO, $\mathrm{H_{2}O}$) and of dust grains (e.g., Forsterite -  Mg$_{2}$SiO$_{4}$, Ensaltite - MgSiO$_{3}$) which either contribute to the atmospheric opacity or deplete the gas phase from elements.   
 
We selected subgrids of synthetic spectra with 1000 K$\leq T_{eff} \leq$ 3000 K, 3.0 $\leq$ log~g $\leq$ 5.5 ($\geq$ 3.5 below 2000 K), and  $[M/H]$=0. An alternative subgrid of the BT-Settl 2013 models (1000 K$\leq T_{eff} \leq$ 3000 K, 3.5 $\leq$ log~g $\leq$ 5.5) with  M/H=+0.5 dex was also used to explore the effect of the metallicity on the determination of log~g and $T_{eff}$.  The spacing of the model grid  is 100~K and 0.5~dex in log~g.

\begin{figure*}[h!]
\centering
	\includegraphics[width=18cm]{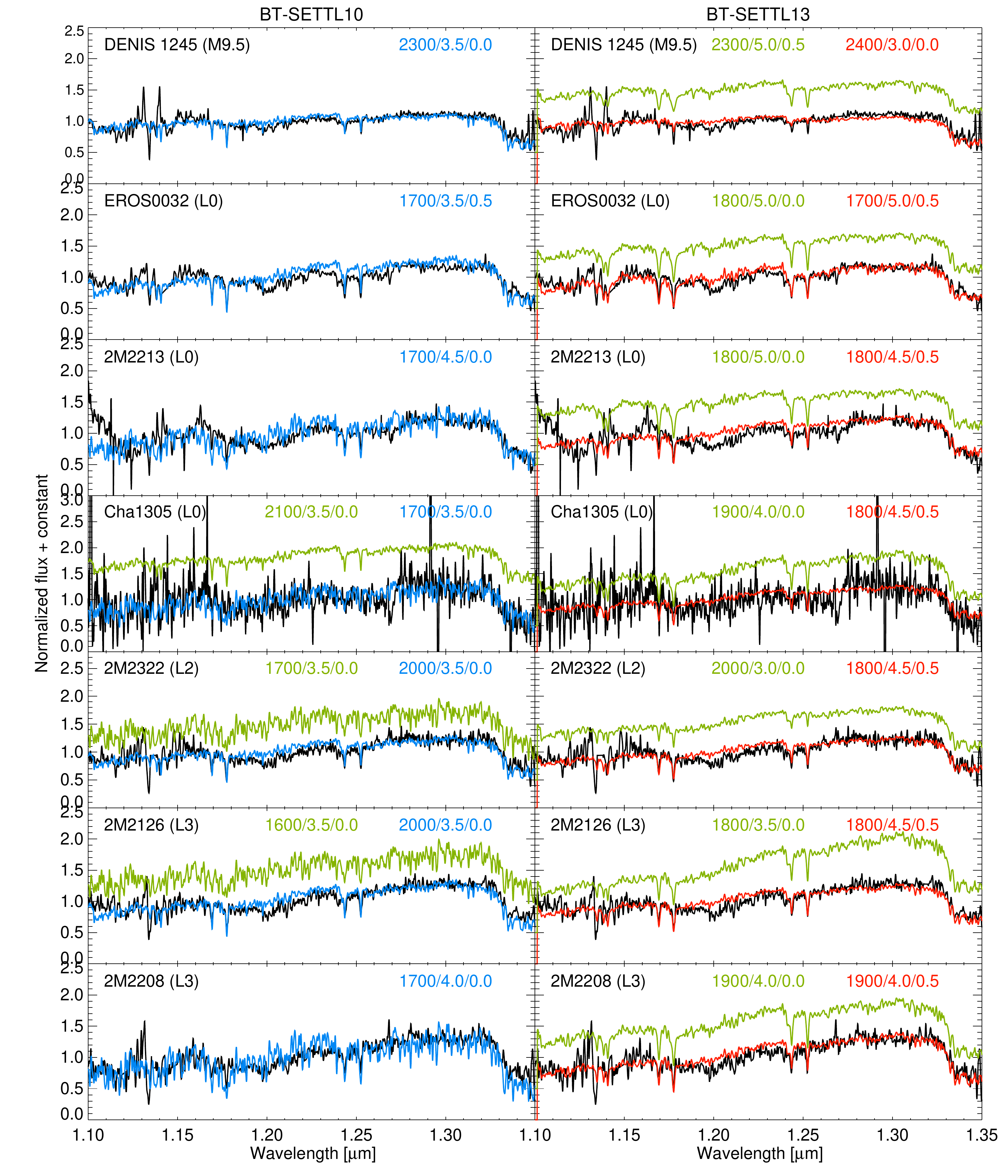}
	\caption{Visual comparison of the best fit BT-Settl 2010 (left panel, blue) and BT-Settl 2013 (right panel, red) synthetic spectra to the six new  ISAAC spectra of young M9.5-L3 dwarfs in the J band (1.1-1.35~$\mu$m). All spectra have been normalized over the wavelength interval. Alternative solutions are shown in light green and are shifted by +0.3 to +0.5 flux normalized units for clarity.}
\label{SynthJ}
\end{figure*}

\begin{figure*}[h!]
\centering
	\includegraphics[width=18cm]{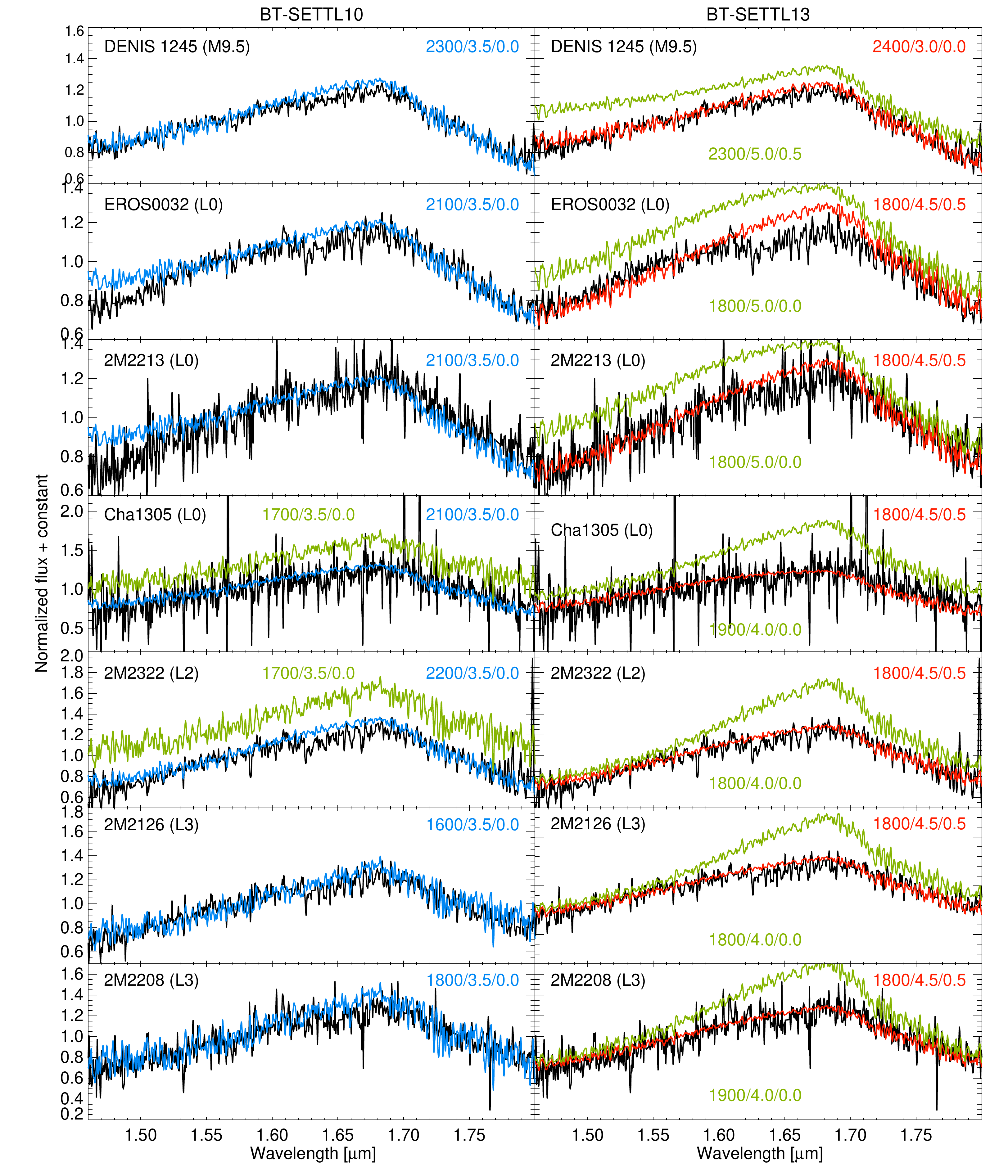}
	\caption{Same as Figure \ref{SynthJ} but in the H band (1.46-1.8 $\mu$m).}
\label{SynthH}
\end{figure*}

\begin{figure*}[h!]
\centering
	\includegraphics[width=18cm]{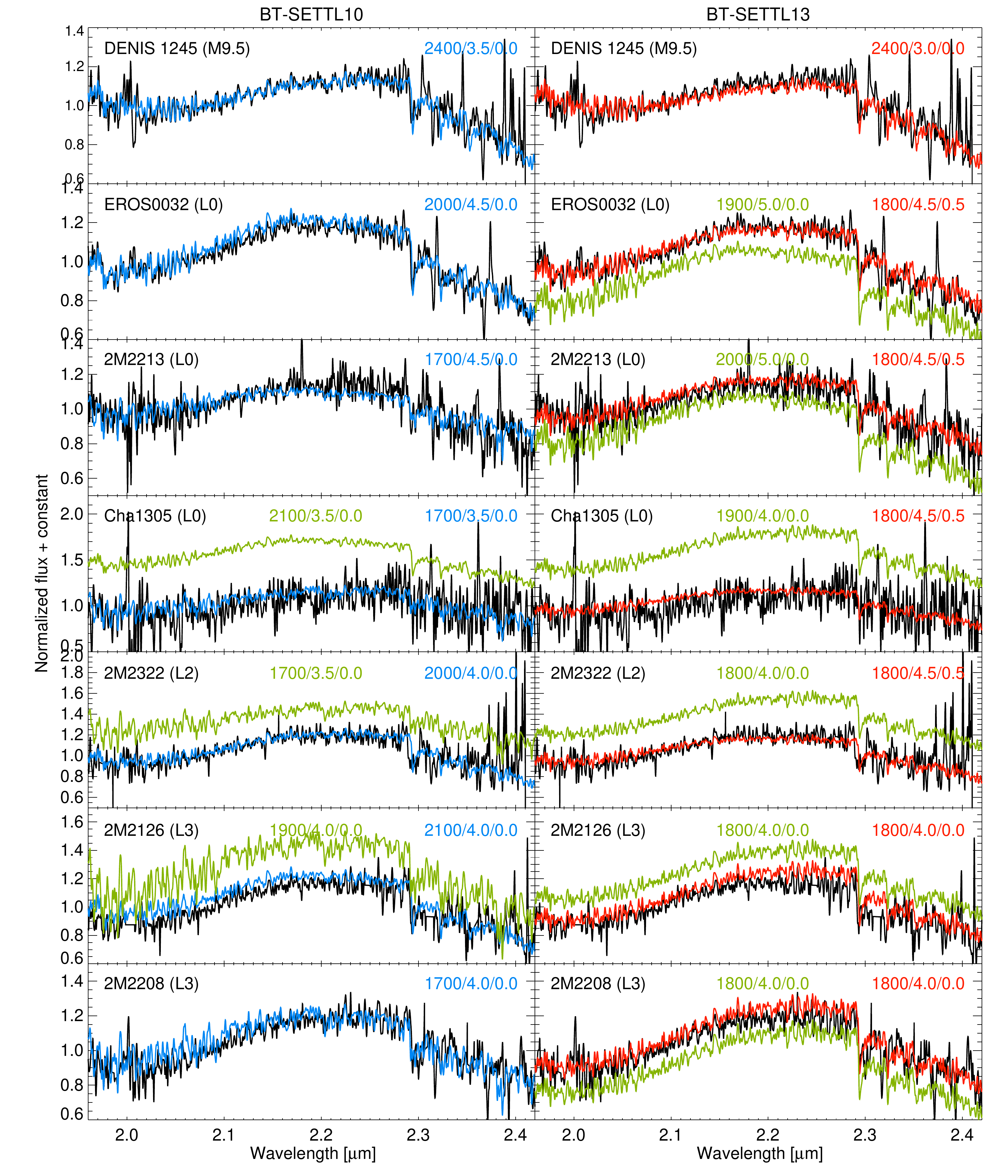}
	\caption{Same as Figure \ref{SynthJ} but in the K band (1.96-2.42 $\mu$m).}
\label{SynthK}
\end{figure*}

\begin{figure*}[h!]
\centering
	\includegraphics[width=18cm]{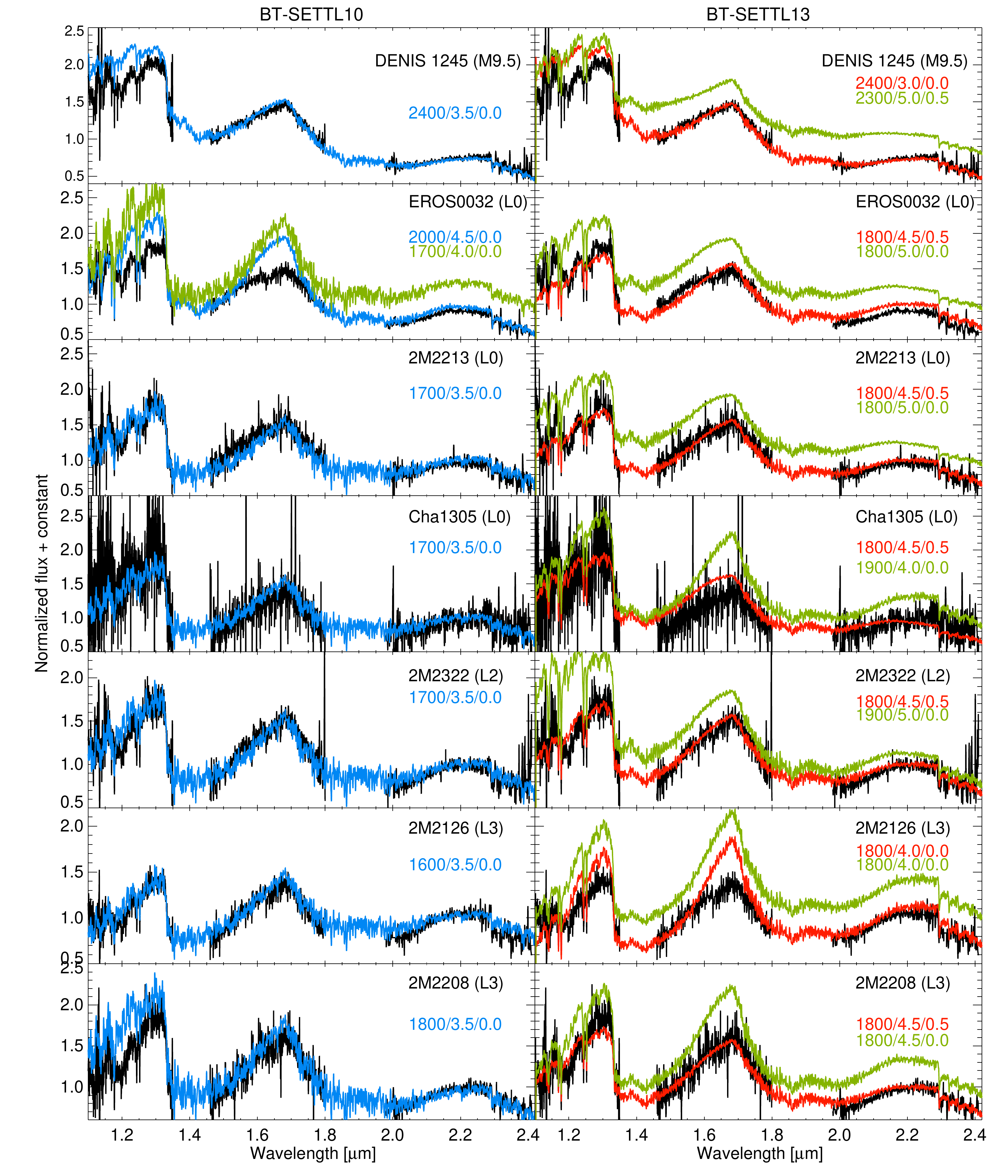}
	\caption{Same as Figure \ref{SynthJ} but for the whole near-infrared spectrum (1.1-2.42 $\mu$m).}
\label{SynthJHK}
\end{figure*}
  
     \subsection{Near-infrared spectra}
     \label{subsub:NIRspec}

BT-Settl 2010 and 2013  synthetic spectra were smoothed to the resolution of ISAAC. The models were then re-interpolated on the ISAAC wavelength grid. Spectra were normalized in the following wavelength intervals 1.1-1.35~$\mu$m (J), 1.46-1.80~$\mu$m (H),  2.02-2.42 $\mu$m (K), and  1.1-2.42 $\mu$m (JHK), and compared in these intervals using least-squares. Results from the fit were always checked visually.  This often revealed innapropriate fitting solutions  induced by remaining uncertainties in the models. Fits with the BT-Settl 2010 model spectra were in particular affected by a numerical noise  introduced by the limited original wavelength sampling of the models \citep{Bonnefoy2013}. The fits with the BT-Settl 2013 models were more affected by the non simultaneous fit of the water bands longward 1.33 $\mu$m and by the  improper modeling of the H~band shape. Atmospheric parameters corresponding to the best fitting models are reported in Table \ref{Tab:atmopar}. $T_{eff}$, $log g$, and $M/H$ have minimum uncertainties of 100 K, 0.5 dex, and 0.5 dex respectively. These errors correspond to the sampling of the atmospheric parameters of the model grids. We show the best fitting synthetic spectra in Figures \ref{SynthJ} to \ref{SynthJHK}. 

BT-Settl models reproduce the shape of the pseudo-continuum and of the prominent narrow atomic (K I, Na I) and molecular (CO) lines of the objects when each band is fitted independently from each other. We recover the non-reproducibility from 1.6 to 1.7 $\mu$m and from 1.195-1.205 $\mu$m quoted in \cite{Bonnefoy2013} and, at least partially, due to missing FeH opacities. These non-reproducibilities get stronger in the case of EROS~J0032. The spectrum of this object has features more typical of mature field dwarfs (Section \ref{empirical_analysis}). This miss-match is  consistent with the conclusions of \cite{Bonnefoy2013} who find a similar departure of the models at these wavelengths in field L-dwarf spectra.  

The surface gravity of Cha1305 and DENIS J1245 found with  BT-Settl 2010 falls in the range expected from evolutionary models for 1-10 Myr objects. It also corresponds, within error bars, to the one determined for late-M/early-L targets from Chameleon and TW Hydrae by  \cite{Bonnefoy2013} based on the same analysis tools and models. The BT-Settl 2013 models fit a higher surface gravity and metallicity to the spectrum of Cha1305. The two parameters are known to have counteracting effects on  the atmospheric pressure at a given optical depth \citep[see][]{2007ApJ...657.1064M} and dust content in the atmosphere. They can then induce opposite variations of the main spectroscopic features \citep{2008ApJ...686..528L}.  Therefore, the higher surface gravity found for Cha1305 is likely caused by the degenerate effect of metallicity. We find an alternative solution with a lower surface gravity for this object at solar-metallicity (see Figure \ref{SynthJHK}), which provides a good fit of the 1.1-2.5 $\mu$m spectrum, but fails to reproduce the shape of the pseudo-continuum in the H band. 

The fit of the J, H, and K band spectra do not reveal a clear correlation between the spectral type and the temperature when the 2010 release of the models is used. The best-fitting temperatures can vary by up to  500 K from band to band. The 1.1-2.5 $\mu$m spectra are also fitted by models at lower temperatures, with differences of 500 K. The spread in effective temperature is reduced to 200 K with the 2013 models. Nevertheless, the solutions found fitting the 1.1-2.5 $\mu$m spectrum of the objects  provides a  reasonable fit in the individual bands with both models.  

The reduced scatter in effective temperatures found with the 2013 release of the models also reflects the overall better quality of the fit provided by these models. BT-Settl 2010 models reproduce the spectra better  than the 2013 models when solar-metallicity models are considered. These behaviors are closely related to the dust allowed to  form and be sustained in the  atmospheres of both models. We discuss these differences in Section~\ref{Subsec:behav}. 

We decided to further test the models by comparing their predictions to the 1-5 $\mu$m spectral energy distributions of the objects.

\begin{table*}
  \caption{Atmospheric parameters corresponding to the best-fit spectra or synthetic fluxes for our seven targets. We give $T_{eff}/log\:g /[M/H]$.}
  \label{Tab:atmopar}
\begin{center}
\begin{tabular}{lllllllll}
 \hline
 \hline
Model   &   Band   &   DENIS J1245   &  EROS~J0032  & 2M2213 &   Cha 1305 &  2M 2322  & 2M 2126  &   2M 2208  \\
 \hline
 BT-Settl 2010  &  J   &  2300/3.5/0.0  &   2100/3.5/0.0  &  1700/4.5/0.0	& 1700/3.5/0.0   &  2000/3.5/0.0  &  2000/3.5/0.0   &  1700/4.0/0.0  \\
BT-Settl 2010 &  H   & 2300/3.5/0.0  & 2100/3.5/0.0    &  1700/4.5/0.0	&2100/3.5/0.0  &  2200/3.5/0.0  &  1600/3.5/0.0  &  1800/3.5/0.0  \\
BT-Settl 2010  &  K   & 2400/3.5/0.0    & 2000/4.5/0.0    & 1700/4.5/0.0 	&1700/3.5/0.0 & 2000/4.0/0.0  &  2100/4.0/0.0  &  1700/4.0/0.0  \\
BT-Settl 2010  &  JHK   &  2400/3.5/0.0  & 2000/4.5/0.0    & 1700/3.5/0.0 	& 1700/3.5/0.0 & 1700/3.5/0.0  &  1600/3.5/0.0  &  1800/3.5/0.0  \\
BT-Settl 2010  &  SED   &  2200/3.5/0.0  & 2000/4.5/0.0  & 2000/3.0/0.0	& \dots    & 1800/3.5/0.5  & 1800/3.5/0.0 &  1800/4.0/0.0  \\
\hline
 BT-Settl 2013  &  J   &  2400/3.0/0.0  &   1700/5.0/0.5  & 1800/4.5/0.5 	&	1800/4.5/0.5   &  1800/4.5/0.5  &  1800/4.5/0.5   &  1900/4.0/0.5  \\
BT-Settl 2013 &  H   & 2400/3.0/0.0  & 1800/4.5/0.5    & 1800/4.5/0.5	& 1800/4.5/0.5  &  1800/4.5/0.5  &  1800/4.5/0.5  &  1800/4.5/0.5  \\
BT-Settl 2013  &  K   & 2400/3.0/0.0    & 1800/4.5/0.5    &  1800/4.5/0.5	&1800/4.5/0.5 & 1800/4.5/0.5  &  1800/4.0/0.0  &  1800/4.0/0.0  \\
BT-Settl 2013  &  JHK   &  2400/3.0/0.0  & 1800/4.5/0.5    &  1800/4.5/0.5	&1800/4.5/0.5 & 1800/4.5/0.5  &  1800/4.0/0.0  &  1800/4.5/0.0  \\
BT-Settl 2013  &  SED   &  2100/3.0/0.0  & 1900/4.0/0.5  & 1800/4.5/0.0	& \dots    & 1800/3.5/0.5  & 1800/3.5/0.0 & 1800/4.0/0.0  \\
\hline
\end{tabular}
\end{center}
\end{table*}

     \subsection{Spectral energy distributions}\label{Spectral_energy_distributions}

We built the spectral energy distributions (SED) of the sources using published photometry from the 2MASS \citep[J, H, K bands; $\lambda_{ref}=1.235$, 1.662, and 2.159 $\mu$m ;][]{2003yCat.2246....0C} and WISE \citep[W1, $\lambda_{ref}=3.4\:\mu$m;  W2, $\lambda_{ref}=4.6\:\mu$m;][]{2012yCat.2311....0C} sky surveys. The sources also have W3 ($\lambda_{ref}=12\:\mu$m) and W4 ($\lambda_{ref}=22\:\mu$m) WISE photometry. Nevertheless, we refrained from accounting for this photometry in the fit since the sources were not detected at a good S/N ($> 8$) in the WISE images. We excluded Cha 1305 from this analysis since the SED of this source has a strong excess \citep{2006ApJ...644..364A}.  The optical photometry, avaliable for some objects, is not included in the fit because the models are known to be inaccurate at these wavelengths \citep[see][]{Bonnefoy2013}.

The infrared photometry was converted to fluxes using tabulated zero points (\citealt{2003AJ....126.1090C, 2011ApJ...735..112J}).

We generated synthetic fluxes from the BT-Settl 2010 and 2013 model grids in the passbands corresponding to the avaliable photometry of the sources. The synthetic fluxes ($F_\lambda$) are provided per square centimeters of the stellar surface. The synthetic SEDs were then normalized to the object apparent fluxes  by scaling a dillution factor which minimized the $\chi^{2}$ of the fit. This dillution factor corresponds to $R^2/d^2$, which $R$  and $d$ the radius and distance of the source, respectively.

\begin{figure*}[h!]
\centering
	\includegraphics[width=\textwidth]{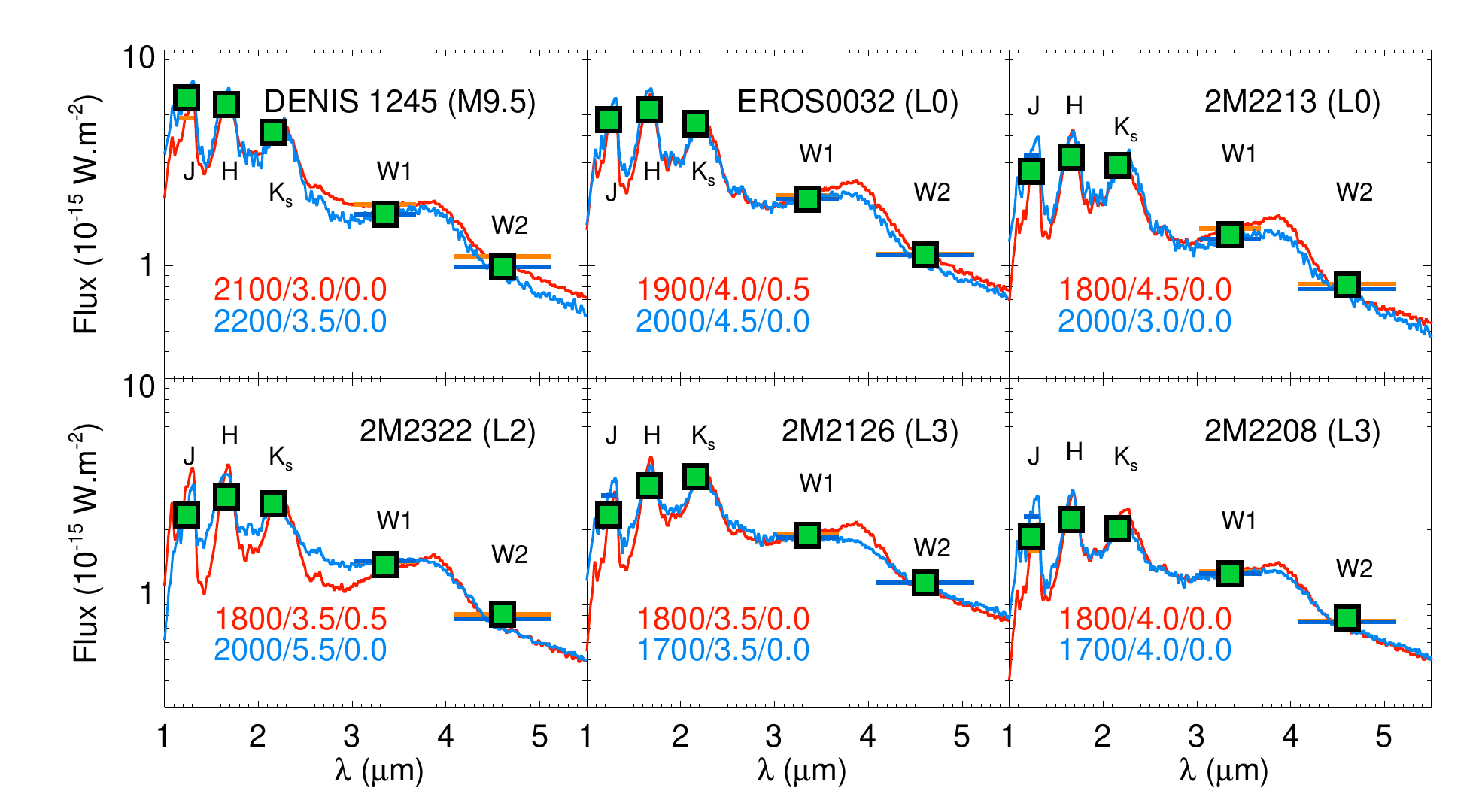}
	\caption{Fit of the spectral energy distribution of five sources of the target sample  (green squares) without noticeable excess emission by BT-Settl 2010 (blue) and BT-Settl 2013 (red) synthetic fluxes (laying bars). The corresponding best fit spectra are overlaid.}
\label{SED}
\end{figure*}

The best fitting atmospheric parameters corresponded to the minimimum $\chi^{2}$ of the fit found for all possible combiations of atmospheric parameters in the grid of models. They are reported in Table \ref{Tab:atmopar}.  The corresponding spectra are shown in Figure \ref{SED}. $\chi^{2}$ maps indicate that the fit is  sensitive to T$_{eff}$ only, for most of the objects. We estimate errors of $^{+300}_{-100}$ K for DENIS~1245, and $\pm$100 K for the remaining targets with the BT-Settl 2013 models. We also find errors of $^{+100}_{-500}$ K for 2M2213 and 2M2322, and $\pm$200 K otherwise for the BT-Settl 2013 models.  These errors are based on 5$\sigma$ contours in the $\chi^{2}$ maps.  Both models yield similar effective temperatures ($\pm$ 200~K). This is  not surprising since differences between the models are expected to vanish at the spectral resolution corresponding to the broad-band filters considered here. This also indicates that the errors are conservative.

Semi-empirical radii $R$ can be derived for the objects with known distances $d$ from the normalization factor $R^{2}/d^{2}$ used to scale the model SED to the flux of the object. This is the case for EROS~J0032 and DENIS~1245 (see Section \ref{subsec:prop}). We therefore derive $R=0.9\pm0.2\:R_{Jup}$ for  EROS~J0032 and $R=2.4\pm0.6\:R_{Jup}$ for DENIS~1245, from the two sets of models. We compare these values to predictions from evolutionary models in Section \ref{subsec:prop}.

\begin{figure}[h!]
\centering
	\includegraphics[width=\columnwidth]{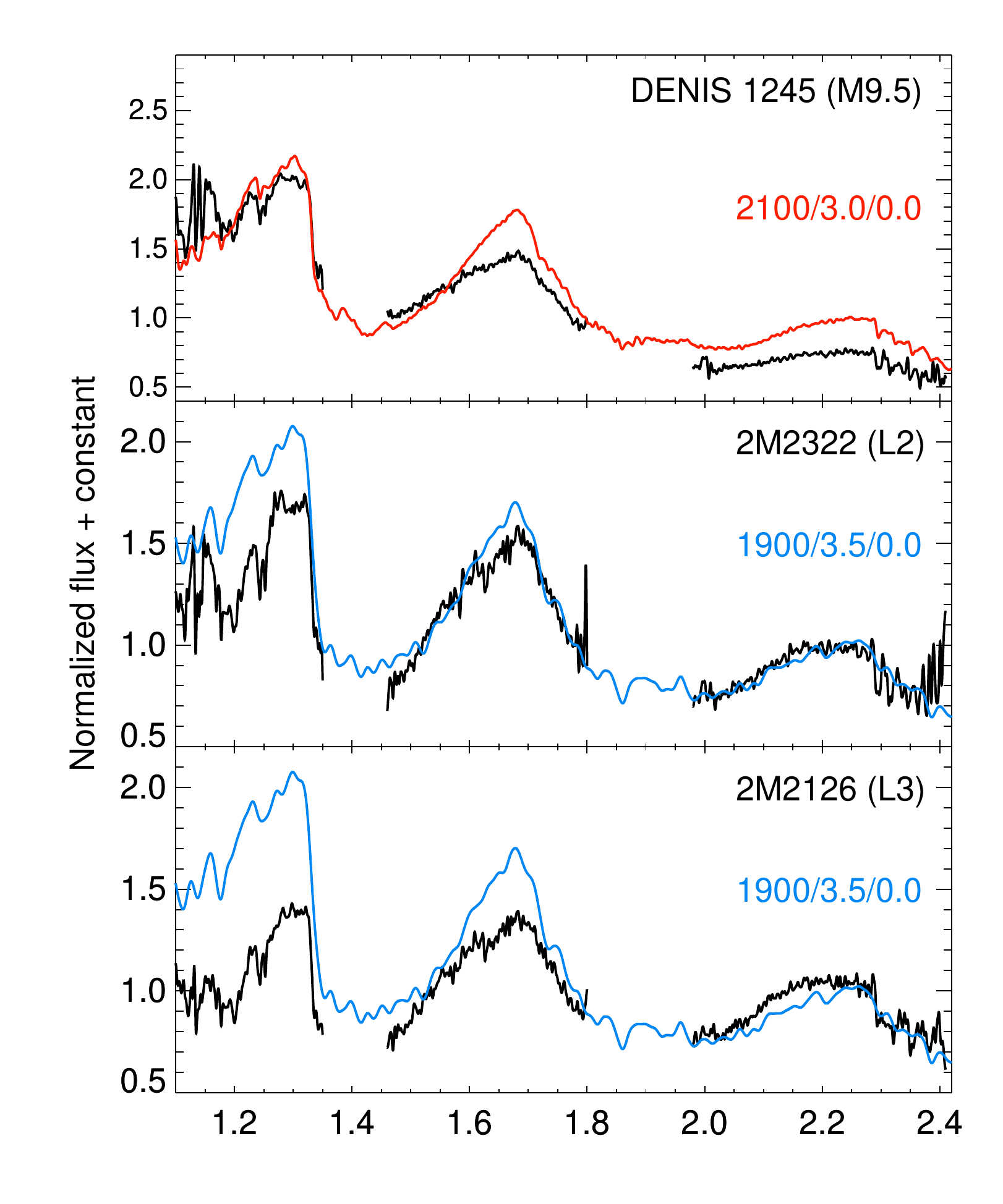}
	\caption{ISAAC spectra (black) smoothed at R=300 compared to best-fitting BT-Settl 2010 (blue) and  for BT-Settl 2013 (red) spectra with previously determined effective temperature inferred from the SED fit for objects for which an independent fit of the 1.1-2.5 $\mu$m does not yield the same effective temperature.}
\label{JHKSED}
\end{figure}

 The SED fit confirms  the effective temperatures found from the near-infrared spectra. Two objects (2M2322, and 2M2126) have best fit temperatures which disagree with the ones found from the fit of the BT-Settl 2010 models to the 1.1-2.5  $\mu$m spectra.  We selected a sub-grid of synthetic spectra corresponding to the set of atmospheric parameters producing the best fit of the SED within a 5 $\sigma$ confidence level. We identified the spectrum from this sub-grid producing the best fit of the 1.1-2.5 $\mu$m spectra of the targets, and displayed them in Figure \ref{JHKSED}.  The best fits are found for low surface gravities, therefore further confirming results from the empirical analysis. Nevertheless, the comparison demonstrates that the models do not successfully reproduce the global spectral slope at these wavelengths as well as the pseudo-continuum in the H~band. We reach similar conclusions for DENIS~1245, whose fit also indicates a mismatch of the overall spectral slope by the models if reduced (3$\sigma$) errors are  considered on the temperature derived from the SED.  We discuss possible explanations for the non reproducibility of the models in the following Section.

\section{Discussion}\label{discussion}

\subsection{Behavior of atmospheric models}
\label{Subsec:behav}
The atmospheric models yield atmospheric parameters that are mostly consistent with our empirical analysis. All the L${\gamma}$ object have surface gravities expected for  young objects. BT-Settl 2013 models converge toward an overall higher surface gravity for EROS~J0032.

 Surprisingly, all spectra but the one of DENIS~1245 are well reproduced by a single BT-Settl 2013 synthetic spectrum with $T_{eff}$=1800 K, log~g $=$ 4.5, and $M/H=+0.5$.  Models at higher metallicity used in this analysis had not been extensively tested, and suffer from
increased numerical instability due to the several times greater dust content in these atmospheres.  Non convergent models can indeed have an anomalous dust content, that can sometimes produce spectra which match well the observations. We found nonetheless that a neighboring spectra with $T_{eff}=1700$ K and log~g $=$4.5 is affected by this problem. Therefore, it is possible that our temperature estimate could be biased by 100 K given  our inability to check if a convergent model at these $T_{eff}$'s can provide a better fit. We also verified that the features of the model spectrum at $T_{eff}=1800$ K are coherent with those found in model spectra for other neighboring $T_{eff}$ and log~g. 
 
The near-infrared spectral slopes, and therefore results from the fits of the ISAAC spectra, are mostly tied to the dust content in the atmosphere. Solutions at high metallicity can then be interpreted as if the 2013 solar metallicity models were not forming enough dust in the atmosphere compared to the 2010 models. This lack of dust could also explain why the models do not reproduce well the shape of the pseudo-continuum in the H~band. The mismatch found in the BT-Settl 2010 models (Figure \ref{JHKSED}) is also indicative of a lack of dust grains at high altitude/low optical depths in the cloud model. The problem may be solved by an ongoing revision of the models. The current version of the code uses a mixing length parameter (which parametrizes the vertical size of the convection cells) of 2 throughout the regime from M dwarfs to brown dwarfs. The new RHD simulations rather indicate that this parameter should be set to lower values. In addition, the grain growth in the 2013 models was artificially suppressed as it was linked to the assumed availability of nucleation seeds. The next version of the models will  treat growth by coagulation in a more consistent fashion and could in principle produce thicker clouds.

The fit of the SED is  less influenced by localized errors in the models because of the extended spectral coverage and the lower resolution of the fit, set by the narrowest filter. If we assume that the SED fits are more reliable, we confirm the quick drop of the effective temperature at the M/L transition discovered for  young optically-classified dwarfs (GSC 08047-00232 B, OTS 44, KPNO-Tau 4) by \cite{Bonnefoy2013}.  The effective temperature remains nearly constant for the L0-L3${\gamma}$ dwarfs of the sample. These temperatures are close to those of other young L-type low-mass companions \object{AB Pic b} \citep{Bonnefoy2010}, 1RXS1609b \citep{2008ApJ...689L.153L, 2010ApJ...719..497L}, \object{GJ 417 B} \citep{Bonnefoy2013}, $\beta$ Pictoris b \citep{2013A&A...555A.107B}, G196-3B \citep{Zapatero-Osorio}, and CD-35 2722B  \citep{2011ApJ...729..139W}. This further suggests that $\beta$ Pictoris b is an early-L dwarf.

\subsection{Revised properties}
\label{subsec:prop}
We combine results from analysis of the SED and the 1.1-2.5 $\mu$m ISAAC spectra to derive final estimates for the objects and report them in Table \ref{Tab:atmoparadopt}. We prioritized the solutions derived  from the SED fit with the BT-Settl 2013 models for the final estimate of the $T_{eff}$. The values of the surface gravity correspond to the most frequent solutions found from the fit of the J, H, K, and JHK~band spectra. The error bars were derived on a case by case basis from the dispersion of the atmospheric parameters.

These new parameters and complementary material found in the literature are used to re-discuss the properties of DENIS J1245, 2M0032,  Cha 1305, and 2M2213.

\begin{table}
  \caption{Adopted atmospheric parameters and bolometric luminosity}
  \label{Tab:atmoparadopt}
\begin{center}
\begin{minipage}{\linewidth}
\begin{tabular}{llll}
 \hline
 \hline
Object   &   $\mathrm{T_{eff}}$ [K]   &    log g  &   log$_{10}$(L/L$_{\odot}$)\tablefootmark{a}  \\
 \hline
 \hline
DENIS J1245  &  $2200\pm200$   &   $3.0\pm0.5$   &   $-3.02 \pm 0.21$ \\     
EROS~J0032  &  $1900 \pm 200$   &   $4.5\pm0.5$   &   $-3.93 \pm 0.11$   \\ 
2M 2203  & $1800 \pm 100$   &   $4.0 \pm 0.5$   & \dots  \\
Cha 1305 &  $1800\pm100$  &   $4.0 \pm 0.5$  &  $-3.13\pm0.08$   \\
2M 2322  & $1800 \pm 100$   &   $4.0 \pm 0.5$  & \dots \\
2M 2126  &   $1800 \pm 100$  &   $4.0 \pm 0.5$ & \dots \\
2M 2208  & $1800 \pm 100$  &  $4.0 \pm 0.5$& \dots  \\
\hline
\end{tabular}
\end{minipage}
\tablefoot{ 
\tablefoottext{a}{Re-computed based on the bolometric corrections of young M9.5 and L0 dwarfs reported in \cite{2010ApJ...714L..84T}.}
}
\end{center}
\end{table}

\begin{table}
  \caption{Physical properties of the objects with known distance and age.}
  \label{Tab:physprop}
\begin{center}
\begin{minipage}{\linewidth}
\begin{tabular}{llll}
 \hline
 \hline
Object   &   Age  &  M   from L/L$_{\odot}$\tablefootmark{a}    &  M  from $\mathrm{T_{eff}}$\tablefootmark{a}  \\
			 &    (Myr)      &   ($\mathrm{M_{Jup}}$)    &  ($\mathrm{M_{Jup}}$)  \\
 \hline
DENIS~1245  &  $10^{+10}_{-7}$    &   $16^{+19}_{-7}$  &  $16^{+9}_{-10}$ \\
Cha1305   &  $4\pm2$ &  $12^{+3}_{-4}$  &   $7\pm2$  \\
EROS~J0032    & 30$^{+20}_{-10}$ ?   &  $12 \pm 2$  &   $13_{-1}^{+15}$ \\
EROS~J0032     &  $120\pm20$ ?   & $30 \pm 6$  &   $33 \pm 7$  \\
EROS~J0032 & 21$^{+4}_{-13}$ & 11$^{+1}_{-4}$ & 13$^{+2}_{-4}$\\
\hline
\end{tabular}
\end{minipage}
\tablefoot{ 
\tablefoottext{a}{Derived from the evolutionary models of \cite{Chabrier}.}
}
\end{center}
\end{table}

\subsubsection{DENIS-P~J124514.1-442907 (TWA 29)}\label{DENIS_J1245}

DENIS-P~J124514.1-442907 was identified as a probable member of the TW Hydrae association \citep{1997Sci...277...67K} by \citet{Looper}. Membership was initially based on estimated distance, sky position relative to known association members, proper-motions, H$\alpha$ emission, and features indicative of low surface gravity in low-resolution (R$\sim$120), near-infrared (0.94-2.5 $\mu$m) and medium-resolution (R$\sim$1800), red-optical spectra of the source.  Looper et al. derived spectral types  M9.5 and M9pec respectively from their optical and near-infrared spectra.  \citet{Manara} derived a comparable range of spectral types (M7.2-L0.8) by calculating spectral indices in an 0.58-2.4$\mu$m medium-resolution ( R$\sim$3500) spectrum of DENIS~J1245. These spectral type estimates are in agreement with our value (Table \ref{Tab:Sptypes}).

The membership of DENIS~J1245 in the TW Hydrae association was revisited by both \cite{2012ApJ...754...39S} and \cite{2013ApJ...762..118W}.  Schneider et al. used revised proper motion measurements to assess the membership in the context of other proposed members and found that membership was highly-likely, despite the lack of a measured parallax at the time.  Weinberger et al. measured a parallax and new proper motions from dedicated, multi-epoch photometry and found that the the overall kinematics and Galactic position of the brown dwarf were consistent with the distribution of other higher-mass members of the association.  

We used the online BANYAN tool of \cite{2013ApJ...762...88M} to calculate a membership probability based on the position, proper motion, and measured parallax.  The Bayesian analysis provides a 95\% probability of the brown dwarf being a member of the TW Hydrae association based only on the available kinematics (i.e. not considering the evidence for youth). We also use the methods of \cite{2009AJ....137.3632L} and \cite{2010AJ....140..119S, 2012AJ....143...80S} to constrain group membership.  We calculated $\phi$, the angle between the source proper motion vector and that expected for the average motion of kinematic moving group members at its position, and d$_{kin}$, the source's distance assuming it is a group member.  We checked these values for each of the young, kinematic groups described in \cite{2008hsf2.book..757T}. We found $\phi$ for DENIS~J1245 was smallest when calculated for both the $\beta$ Pictoris moving group and the TW Hydrae association, $\sim$5.5$^{\circ}$.  This angle is typically $\lesssim$10$^{\circ}$ for well established group members.  The d$_{kin}$ predicted for both moving groups was also very similar, $\sim$95 pc, and generally consistent with the 79$\pm$13 pc distance measured by \cite{2013ApJ...762..118W}.  However, the Galactic XYZ distances of the brown dwarf are most consistent with other members the TW Hydrae association, particularly the positive Z distance.  These results are consistent with previous kinematic studies, and when combined with previous evidence for youth and the spectral features indicative of very-low surface gravity in our ISAAC spectrum, indicate DENIS~J1245 is a very strong candidate for TWA membership. Designation as a true member will require an accurate RV measurement.  

 \citet{Witte} compared their  DRIFT-\texttt{PHOENIX} model spectra to the near-infrared spectrum of \cite{Looper} to derive $T_{eff}$=1900~K, log g $=$ 4.5 and $[M/H]= 0.0$. Our estimates of these physical parameters are consistent with these values if we consider a 100 K error on their measurements, corresponding to the sampling in effective temperature of their model template grid. Nevertheless, they disagree with the temperature derived from the fit of the 1.1-2.5 $\mu$m only. Our estimates of the effective temperature (from the SED, or the spectra) are  in good agreement with the temperature ($T_{eff}=2300$ K) derived from the  extension \citep{2002ApJ...580..317B} of the spectral type temperature conversion scale of \cite{1999ApJ...525..466L} and \cite{2003ApJ...593.1093L}. We derive a mass of $16^{+9}_{-10} \mathrm{M_{Jup}}$ for DENIS~J1245 by comparing our  temperature (reported in Table \ref{Tab:atmoparadopt}) to predictions of the DUSTY evolutionary models \citep{Chabrier} for an age of $10^{+10}_{-7}$ Myr \citep{2006A&A...459..511B}. 

We used the parallax measurement of \cite{Weinberger} and the $BC_{K}$ for young M9.5 dwarfs derived by \cite{2010ApJ...714L..84T} to estimate the bolometric luminosity (Table \ref{Tab:physprop}) of the source.  This luminosity corresponds to a predicted mass consistent with that derived from $T_{eff}$ (see Table \ref{Tab:physprop}). We thus confirm that to date, DENIS~J1245 is the lowest mass isolated object proposed to be a member of the TW Hydrae association. The semi-empirical radius derived in Section \ref{Spectral_energy_distributions} is consistent with evolutionary models predictions \citep{Chabrier} for the estimated age of the object. 


\subsubsection{EROS-MP~J0032-4405}\label{2M0032}

\begin{figure*}[t]
\centering
	\includegraphics[width=18cm]{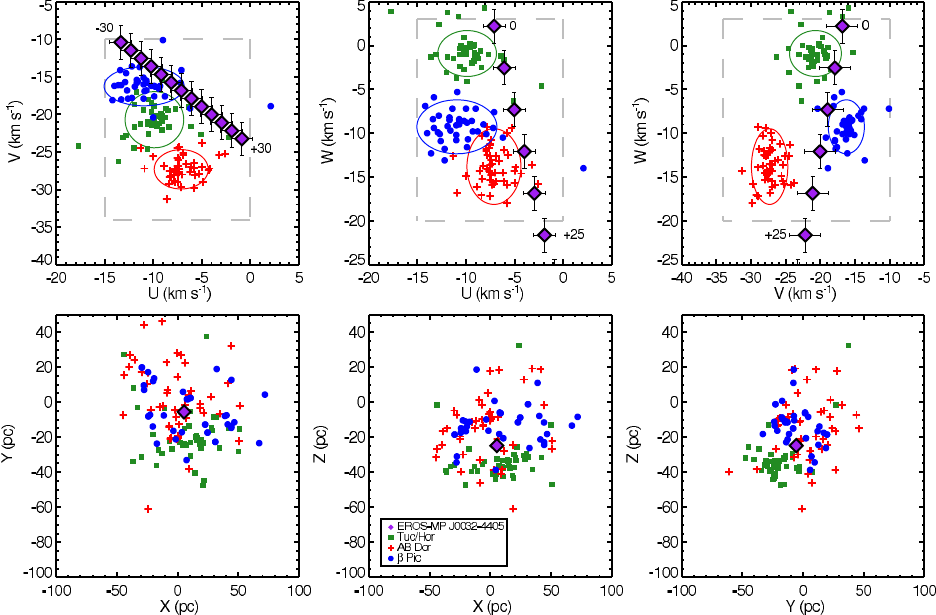}
	\caption{Projections of the 6D Galactic kinematics of EROS~J0032 compared to the young Tucana/Horologium, AB Doradus, and $\beta$ Pictoris kinematic groups.  EROS~J0032 is presented by a violet diamond with associated error bars. The multiple points represent different possible values of the radial velocity (RV) ranging from -30 km s$^{-1}$ to +30 km s$^{-1}$ in increments of 5 km s$^{-1}$.  The young kinematic groups are designated in the figure legend. Top panels: Projections in ($UVW$) Galactic velocity.  The colored ellipses designate the 2$\sigma$ dispersion of the average velocities of each group from \cite{2013ApJ...762...88M}. EROS~J0032 is most consistent with the Tuc/Hor and $\beta$ Pic groups for RVs between $\sim$5 and 10 km s$^{-1}$.  Bottom panels: Projections of the ($XYZ$) Galactic distance. EROS~J0032 is consistent with the distributions of all three young groups. For reference, we also plot the Galactic velocity space occupied by most young stars in the solar neighborhood as a dashed, grey box \citep{2004ARA&A..42..685Z}.}
\label{6Dkin}
\end{figure*}

{ EROS-MP~J0032-4405 was discovered by  \citet{Goldman}. \citet{Martin1999} adopted a spectral type of L0 from analysis of an optical spectrum and identified strong Li 6708~\AA~absorption.  The detection of Li constrained the mass to M $\leq$ 50$ M_{Jup}$ and the age to be younger than $\sim$0.5~Gyr. \citet{Goldman} used NG-Dusty models to estimate a temperature of ${\rm T_{eff}}$ = 1850 $\pm$150~K, comparable to our estimate from SED fitting. \citet{Cruz} classified this object as very-low gravity L0$\gamma$ type and estimated a spectroscopic distance of $d_{sp}$=41$\pm$5~pc. \citet{Allers2013} also presented a low resolution near-infrared spectrum and used their index based methods to classify it as L0 VL-G (very-low gravity). Accurate proper motions and a parallax of 38.4 $\pm$ 4.8 mas were presented in \cite{2012ApJ...752...56F}.

Our index based analysis of our ISAAC spectrum provided a spectral type consistent with the L0 determined by \cite{Cruz} and \cite{Allers2013}.  However, the gravity sensitive spectral features in our medium resolution spectrum provide a slightly different view of the brown dwarf's age. Figures 5 and 6 show that the gravity sensitive K I lines and FeH band in the $J$-band and the shape of the $H$-band continuum are consistent with a surface gravity only slightly weaker than typical field brown dwarfs of similar spectra type.  These features are more suggestive of an intermediate surface gravity and thus an older age than the previous optical and near infrared spectra. Since our knowledge of how dust and low-surface gravity affect individual spectral features and overall spectral morphology at different wavelengths and spectral resolutions is incomplete, it is difficult to assess this discrepancy.  As shown for two proposed L-type members of the AB Doradus moving group in \cite{Allers2013}, even the spectra of brown dwarfs at the same purported age and temperature exhibit features consistent with different surface gravities.

 \cite{Cruz} showed that the sky position of EROS~J0032 and many other young, field dwarfs were coincident with known members of young kinematic groups.  To follow-up these preliminary suggestions, we followed the same procedures as described for DENIS~J1245 to investigate possible membership of EROS~J0032 in kinematic moving groups. Using only the available proper motions and parallax, the \cite{2013ApJ...762...88M} online tool provides $\sim$92\% membership probability in the Tucana/Horologium association (Tuc/Hor), $\sim$8\% membership probability in the AB Doradus moving group (AB Dor), and <1\% probability in the $\beta$ Pictoris moving group ($\beta$ Pic) and the field. However, the potential membership EROS~J0032 in a young moving group hinges on its radial velocity (RV). 
 
We illustrate this in Fig.~\ref{6Dkin} where we show projections of the 6-dimensional Galactic kinematics\footnote{\emph{UVWXYZ}, where $U$, $V$, and $W$ are Galactic velocities and $X$, $Y$, and $Z$ are Galactic distances.  $U$ and $X$ are positive toward the Galactic center, $V$ and $Y$ are positive in the direction of the Sun's motion around the Galaxy, and $W$ and $Z$ are positive toward the north Galactic pole.  The sun lies at ($UVWXYZ$) = (0,0,0,0,0,0).} of EROS~J0032 for a range of possible RVs ([-30, 30] km s$^{-1}$). The figure reveals that the kinematics of the brown dwarf are a relatively good match to both the Tuc/Hor and $\beta$ Pic groups for RVs between $\sim$ 5-10 km s$^{-1}$. EROS~J0032 is also close to the AB Dor group distribution for RVs $\sim$15-20 km s$^{-1}$, but remains a $>$4$\sigma$ outlier in the ($UV$) plane. EROS~J0032 is consistent with the ($XYZ$) of any of these groups.  To complete the kinematic analysis, we also calculated $\phi$ =  5.3$^{\circ}$, $\phi$ = 0.9$^{\circ}$, $\phi$ = 19.5$^{\circ}$ when comparing the brown dwarf's proper motions to the Tuc/Hor, AB Dor, and $\beta$ Pic groups respectively. The d$_{kin}$'s calculated for the groups are also very comparable; 32 pc for Tuc/Hor, 37 pc for AB Dor, and 26 pc for $\beta$ Pic.  All of these distances are broadly consistent with the measured parallax distance of 26.0 $\pm$ 3.3 pc, but the match to $\beta$ Pic is the best.  Thus, we conclude that the kinematics of EROS~J0032 are most suggestive of possible membership in either the Tuc/Hor or $\beta$ Pic groups, but without a measured RV, definitive membership cannot be assigned.

For completeness, we estimate the mass of EROS~J0032 by comparing our derived effective temperature to DUSTY evolutionary model predictions for two different age ranges: 20 - 40 Myr for possible Tuc/Hor or $\beta$ Pic membership and 130$\pm$20 Myr for possible AB Dor membership \citep{2001ApJ...559..388Z, 2013ApJ...766....6B}}. If the brown dwarf is a member of $\beta$ Pic, its mass is comparable to those of directly imaged planets {(see Table~\ref{Tab:physprop})}. 


\subsubsection{Cha~J130540.8-773958}\label{Cha1305}

 Our new effective temperature determination for Cha~J1305 falls in the same range as the temperatures of 1-3 Myr old M9.5-L0 objects OTS 44 \citep{2004ApJ...617..565L}, KPNO Tau 4 \citep{2002ApJ...580..317B}, and Cha J110913-773444 \citep{2005ApJ...635L..93L}. This is consistent with the observed similarities between the spectrum of OTS 44 and the low resolution (R$\sim$300) near-infrared spectrum of Cha 1305 presented by \cite{Allers2007}.  The optical spectral type (L$0\pm2$) derived by \citet{Jayawardhana} is also consistent with the near-infrared spectral type. 

We recomputed the luminosity of Cha 1305 and report it in Table \ref{Tab:physprop}. This luminosity is based on the $BC_{K}$ of young L0 dwarfs \citep{2010ApJ...714L..84T}, the K~band magnitude of the source, corrected for an $A_{V}=3$ mag \citep{Allers2007, Spezzi}, and an associated distance of  $178\pm18$ pc for the Chameleon II star forming cloud  \citep{1997A&A...327.1194W}.  We compared the temperature and luminosity of Cha~J1305 to \cite{Chabrier} evolutionary tracks and retrieved a mass of 5-15 $\mathrm{M_{Jup}}$ for a Chameleon II age of $4\pm2$ Myr \citep{Spezzi}. This mass is consistent with the previous the mass estimate presented by \citet{Allers2007}.

\subsubsection{2MASS~J22134491-2136079}\label{2M2213}

2MASS~J22134491-2136079 was identified as a peculiar L0 type object with probable low gravity features \citep{Cruz2007}. This gravity classification was based on the strength of VO bands and alkali doublets in the optical.  H$\alpha$ emission was however not detected. \citet{Kirkpatrick2008} estimated an age of less than $\sim$100~Myr via visual comparison to the optical spectra of other brown dwarfs with known ages. \citet{Kirkpatrick2008} considered the estimated age and sky position of    2M~2213  to suggest that it is a possible member of the $\beta$ Pictoris moving group. \citet{Cruz} estimated a distance of $54\pm7$~pc using the $M_{J}$-Spectral type relation from \citet{Cruz2003}. \citet{Allers2013} classified 2M~2213 as spectral type L0 with very-low surface gravity from a low resolution (R$\sim$100) IRTF/SpeX spectrum.  This classification is at odds with the L2pec near-infrared spectral type of \cite{2013arXiv1309.6525M}, likely because \cite{2013arXiv1309.6525M} only compare the spectrum of the source to those of mature field dwarfs.

We report here the first estimate of the temperature of the object using atmospheric models, and find atmospheric parameters which confirm the low surface gravity of the object. To further investigate possible young kinematic group membership, we use the proper motions measured by \cite{2009AJ....137....1F} to apply the same tests of group membership used for DENIS~1245 and EROS~J0032. The BANYAN Bayesian analysis tool provides ambiguous probabilities for group membership: $\sim$30\% $\beta$ Pictoris, $\sim$20\% Tuc/Hor, and $\sim$50\% field.  The angle $\phi$ is also $<$10$^{\circ}$ for many of the kinematic groups discussed in \cite{2008hsf2.book..757T}.  The photometric distance estimated by \citet{Cruz} could be used to potentially rule out membership in several groups, but a parallax measurement is preferred given the young age of 2M 2213.  Thus, the currently available data do not allow for constraints on 2M 2213's possible kinematic group membership. The topic should be revisited once RV and parallax measurements are available for this young brown dwarf.

\section{Conclusions}\label{conclusions}
We obtained and analyzed seven VLT/ISAAC medium-resolution (R$\sim$1500-1700) spectra of M9.5-L3 dwarfs classified at optical wavelengths and showing indications of low surface gravity. We built an age-sequence of M9.5 objects that allow us to pinpoint age-sensitive and gravity-sensitive features at medium-resolving powers. The comparison of our spectra to those of young reference brown dwarfs and companions, and of mature field dwarfs confirm that our objects have peculiar features in the near-infrared indicative  of low surface gravities and young ages. We also confirm the youth of our objects by calculating the equivalent widths of their KI lines and comparing these values per spectral type with the values obtained for young reference brown dwarfs and companions and  mature field dwarfs. We derived near-infrared spectral types based on dedicated spectral indices. These spectral types are in agreement with the optical classification, and confirm the coherence of the classification method.  The analysis revealed that the L2${\gamma}$ object \object{2MASS~J2322} provides a good match to the spectrum of the young planetary mass companion \object{1RXS J160929.1-210524b}.

The spectra and SEDs of the objects can be reproduced by  the 2010 and 2013 BT-Settl atmospheric models.  The 2013 release of the models fits simultaneously the spectra and the SED for the same temperatures at all wavelengths. L0-L3$\gamma$ dwarfs have nearly equal  temperatures around 1800 K. Nevertheless, we  identify that:
\begin{itemize}
\item the 2010 models do not  reproduce the 1.1-2.5 $\mu$m spectral slope of some L2-L3 objects. 
\item the H~band shape is not well reproduced by the BT-Settl 2013 models at solar metallicity. The problem disappears when new, but not as well-tested, models at super-solar metallicity are used, but these models remain mostly untested.
\end{itemize}
Currently, all these discrepancies point out a lack of dust in the cloud models. The next version of the BT-Settl models will modify the treatment of the vertical mixing and of grain growth processes. These new models are expected to produce thicker clouds, and may solve the issues revealed by the ISAAC spectra.

The spectra of the objects will help to  confirm the membership of photometrically-selected candidates  in star-forming regions. Within the next few years, surveys on the next generation of planet imaging instruments such as SPHERE (Spectro-Polarimetric High-contrast Exoplanet REsearch) at VLT, GPI (Gemini Planet Imager) at Gemini South, ScEXAO (Subaru Coronagraphic Extreme AO Project) at Subaru, and LMIRCam (Large Binocular Telescope mid-infrared camera) at LBT should  provide a sample of a few dozen  young companions. Several planets similar to $\beta$ Pictoris b should be unearthed and fall in the same temperature range as our objects. Therefore, our spectra will serve as precious benchmarks for the characterization of the physical and atmospheric properties of these companions. 


\begin{acknowledgements}
We are grateful to the ESO staff for their support during the preparation of the observations and their execution. We acknowledge Emily Rice, Kathlyn Allers, Quinn Konopacky, Stan Metchev, David Lafreni\`{e}re, Jenny Patience, Michael Liu, Tobias Schmidt, Andreas Seifahrt, Brendan Bowler, Zahed Wahhaj, Laird Close, Davy Kirkpatrick, Nadya Gorlova, Carlo Felice Manara, and Catherine Slesnick for providing the spectra of their low-gravity/young objects. We also thank Michael Cushing, John Rayner, and Ian Mc Lean for providing an online access to their spectral librairies.  This research has benefitted from the SpeX Prism Spectral Libraries, maintained by Adam Burgasser at http://pono.ucsd.edu/$\sim$adam/browndwarfs/spexprism. It has also made use of the SIMBAD database operated at CDS, Strasbourg, France. Ga\"{e}l Chauvin, Anne-Marie Lagrange, France Allard, and Derek Homeier acknowledge financial support from the French National Research Agency (ANR) through project grant ANR10-BLANC0504-01.  This project is also supported by the ``Programme  National de Physique Stellaire'' (PNPS) of CNRS (INSU) in France, the European Research Council under the European Community's Seventh  Framework Program (FP7/2007-2013 Grant Agreement no. 247060), and the Lyon Institute of Origins under  grant ANR-10-LABX-66.
This work was supported by Sonderforschungsbereich SFB 881 ``The Milky Way System'' (subproject B6) of the German Research Foundation (DFG).
N. Lodieu was funded by the Ram\'on y Cajal fellowship number 08-303-01-02\@.
This research has been partially supported by the Spanish Ministry of Economics and Competitiveness under the project AYA2010-19136. Fondecyt \#1120299 and Basal initiative PFB06.

\end{acknowledgements}

\bibliographystyle{aa}
\bibliography{young_bds}

\newpage

\begin{appendix}

\section{Properties of the spectra of young companions}
\label{Appendix:A}

We summarize in Table \ref{Table:Companionspec} the main characteristics of the spectra of young companions found in the literature and used in our empirical analysis. We did not reported proposed spectral classes for these companions given the inhomogeneous classification scheme adopted in the literature.

\begin{table*}[!h]
\begin{minipage}{18cm}
  \caption{Characteristics of the young companions spectra}
  \label{Table:Companionspec}
  \centering
\begin{tabular}{llllll}
 \hline
 \hline
Name & Age   &    $\lambda_{min}$  & $\lambda_{max}$ & R & Reference\footnote{References: 1, \cite{Bonnefoy2013}; 2, \cite{Bonnefoy2010}; 3, \cite{2012ApJ...753..142B}; 4, \cite{2013arXiv1307.2237B}; 5, \cite{2011ApJ...743..148B}; 6, \cite{2006ApJ...651.1166M}; 7, \cite{2008ApJ...689L.153L}; 8, \cite{2010ApJ...719..497L}; 9, \cite{Patience}; 10,  \cite{2011ApJ...729..139W}; 11, \cite{Allers2013}; 12, \cite{2007A&A...463..309S}; 13, \cite{2009ApJ...704.1098L}; 14, \cite{Luhman}; 15, \cite{Barman}; 16, \cite{2013Sci...339.1398K}; 17, \cite{Allers2010}; 18,  \cite{2007ApJ...665..736C}.} \\
          &   (Myr)       &   $\mathrm{(\mu m)}$  &  $\mathrm{(\mu m)}$   &   &     \\  
 \hline
2MASS J12073346-3932539 A  &   8   &  1.1   &  2.5   &  1500-2000  &  1   \\
 AB Pic b   &  30   &  1.1   &  2.5   & 1500-2000   &  1, 2 \\
 CT Cha b   &  1-3   &  1.1   &   2.5   &  1500-2000  &  1   \\
 DH Tau b   &  1-3   &  1.1   &   2.5   &  1500-2000  &  1   \\
 Gl 417 B   &  80-250   &  1.1   &   2.5   &  1500-2000  &  1   \\
GSC 08047-00232 B  &  30   &  1.1  &   2.5   &  1500-2000  &  1   \\
HR7329 B  &   12   &  1.1  &   2.5   &  1500-2000  &  1   \\
TWA 5B    &   8   &  1.1   &  2.5   &  1500-2000  &  1   \\
TWA 22A &      12   &  1.1  &   2.5   &  1500-2000  &  1   \\
TWA 22B &      12   &  1.1  &   2.5   &  1500-2000  &  1   \\
USCO CTIO 108B  &  3-11  & 1.1  &   2.5   &  1500-2000  &  1   \\
1RXS J235133.3+312720 B  &  50-150  & 0.8 & 2.5 &  250-1200  &  3  \\
2MASS J01225093-2439505 B  &  10-120   &  1.48  &  2.38  &  3800  &  4 \\
GSC 06214-00210 b   &  3-11  &  1.1  &  1.8   &  3800 &  5 \\
HD203030 b  &  130-400  & 2.0  &  2.6   & 2000  &  6 \\   
1RXS J160929.1-210524 b  &  3-11   & 1.15  & 2.4  &   6000-1300  &  7, 8  \\
2MASS J12073346-3932539 b   &  8  &  1.1  &  2.5   &  1500-2000  &  9  \\
CD-35 2722 B  &  100  &  1.15  &  2.40  &  5000-6000  &  10  \\
G196-3B   &  100  &  1.1  &  2.4  &  2000  &  11  \\
GQ Lup b   &  1-3  &  1.1  &  2.5  &  2000-4000   &  12  \\
GQ Lup b &  1-3  &  1.164  &  2.4  &  5000   &  13  \\
HN Peg b   &  100-500  & 0.65  &  2.56 & 75   &  14  \\
HR8799 b  &  30   &  1.48  &  2.36  &  60  &  15  \\  
HR8799 c  &  30  &  1.965  &  2.381   &  4000  &  16  \\
SDSSJ224953.47+004404.6AB   &  100  & 0.8  &  2.5  & 150  &  17  \\ 
TWA 8B   &  8   &   0.8  &  2.42   &   2000  &  11 \\
TWA 11C  &  8   & 0.95  &  2.42  &  2000  & 11  \\
AB Dor C  &  75-175   &   1.48   &  2.5  &  1500   &  18  \\  
 \hline
\end{tabular}
\end{minipage}
\end{table*}


\section{Best matches of EROS~J0032, 2M~2213, 2M~2126 and 2M~2208 with library spectra.}
\label{Appendix:B}

We present the result of the best matches after performing the empirical comparison of EROS~J0032, 2M~2213, 2M~2126 and 2M~2208 spectra with library spectra in Section \ref{subsec:Lfield?}.

\begin{figure}[!h]
\centering
	\includegraphics[width=0.51\textwidth]{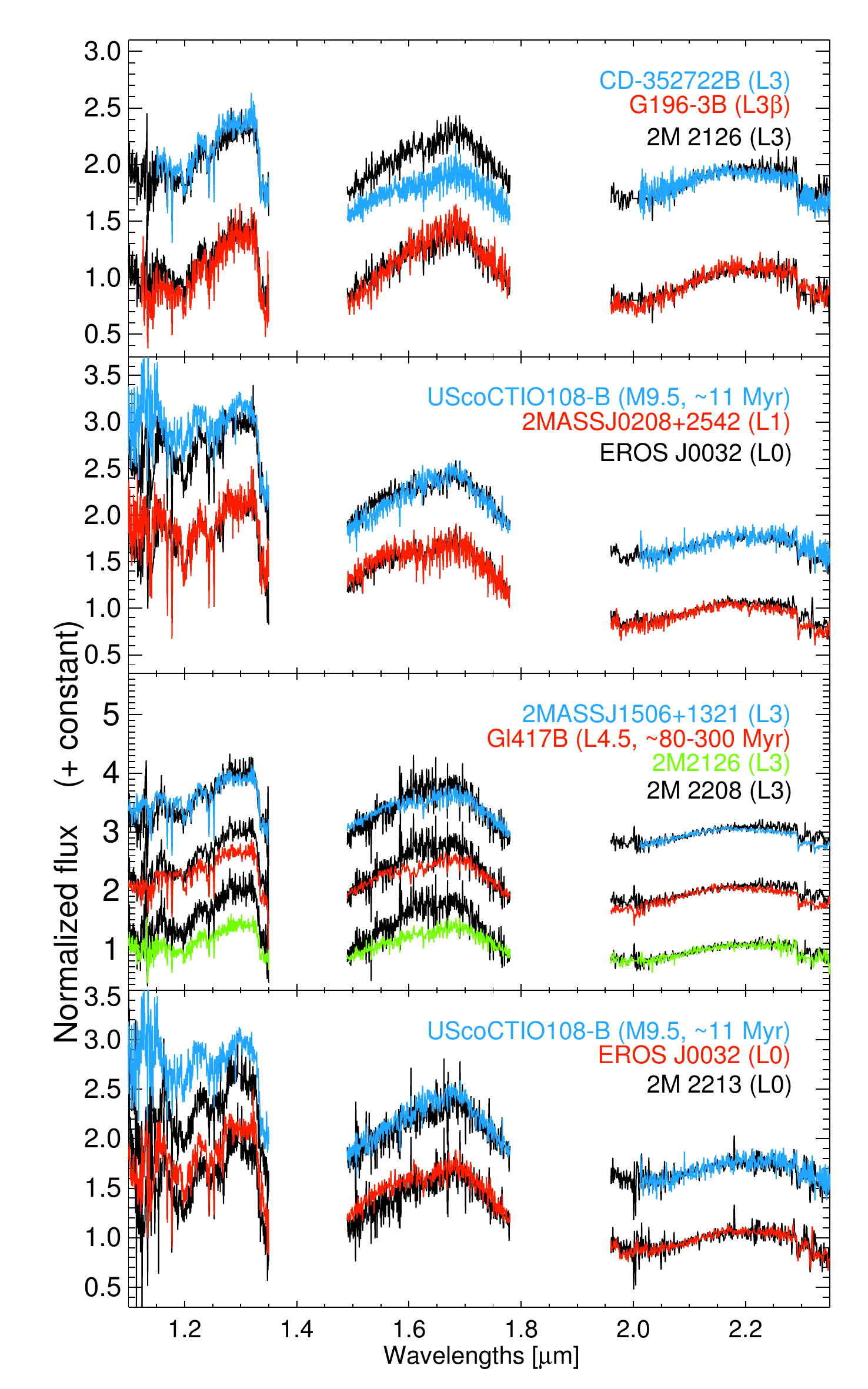}
	\caption{Best matches of the 1.1-2.38~$\mu$m spectra of EROS~J0032, 2M~2213, 2M~2126 and 2M~2208 with library spectra.}

\end{figure}

\begin{figure}[!h]
\centering
	\includegraphics[width=0.51\textwidth]{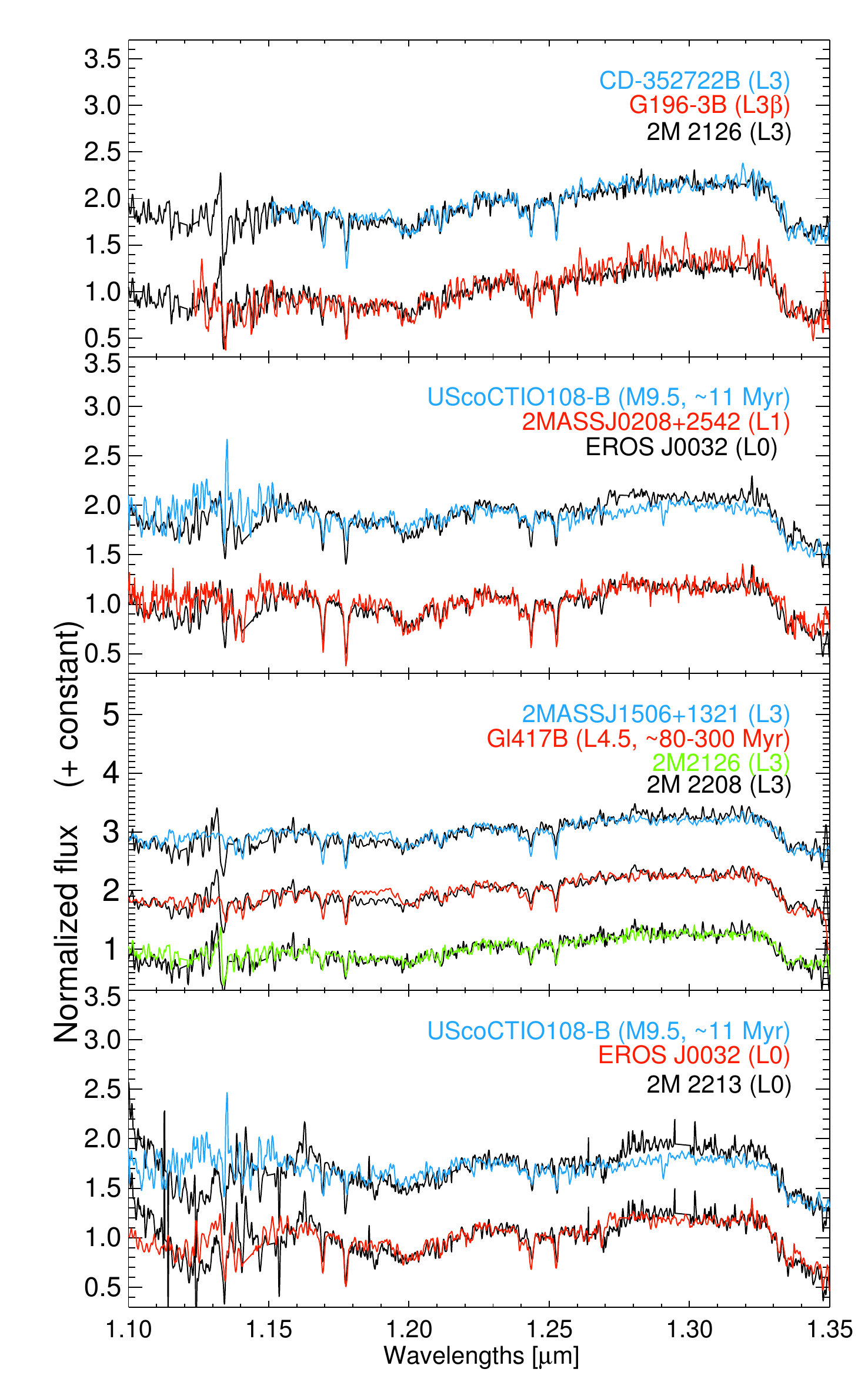}
	\caption{J~band of best matches of the spectra of EROS~J0032, 2M~2213, 2M~2126 and 2M~2208 with library spectra.}
\end{figure}

\end{appendix}

\end{document}